\documentclass[fleqn,usenatbib]{mnras}

\usepackage{newtxtext,newtxmath}

\usepackage[T1]{fontenc}

\DeclareRobustCommand{\VAN}[3]{#2}
\let\VANthebibliography\thebibliography
\def\thebibliography{\DeclareRobustCommand{\VAN}[3]{##3}\VANthebibliography}

\usepackage[normalem]{ulem}  
\newcommand\redout{\bgroup\markoverwith
{\textcolor{red}{\rule[0.5ex]{2pt}{0.8pt}}}\ULon}

\usepackage{graphicx}	
\usepackage{amsmath}	

\usepackage{longtable}
\usepackage{lscape}
\usepackage{multirow}

\usepackage{tikz,xcolor,hyperref}

\definecolor{lime}{HTML}{A6CE39}
\DeclareRobustCommand{\orcidicon}{
	\begin{tikzpicture}
	\draw[lime, fill=lime] (0,0) 
	circle [radius=0.16] 
	node[white] {{\fontfamily{qag}\selectfont \tiny ID}};
	\draw[white, fill=white] (-0.0625,0.095) 
	circle [radius=0.007];
	\end{tikzpicture}
	\hspace{-2mm}
}

\foreach \x in {A, ..., Z}{\expandafter\xdef\csname orcid\x\endcsname{\noexpand\href{https://orcid.org/\csname orcidauthor\x\endcsname}
			{\noexpand\orcidicon}}
}

\title[Low-ionization structures in planetary nebulae]{Low-ionization structures in planetary nebulae - II. Densities, temperatures, abundances and excitation of 6 PNe}
\author[Mari, Gon\c{c}alves and Akras]{
M. Bel\'en Mari$^{1\orcidB{}}$ \thanks{E-mail: mbmari@astro.ufrj.br},
Denise R. Gon\c{c}alves$^{1}$ and 
Stavros Akras$^{2\orcidA{}}$
\\
$^{1}$Observat\'orio do Valongo, Universidade Federal do Rio de Janeiro, Ladeira Pedro Antonio 43, Rio de Janeiro 20080-090, Brazil\\
$^{2}$Institute for Astronomy, Astrophysics, Space Applications and Remote Sensing, National Observatory of Athens, Penteli GR 15236, Greece
}

\date{Accepted XXX. Received YYY; in original form ZZZ}

\pubyear{2022}

\begin{document}
\label{firstpage}
\pagerange{\pageref{firstpage}--\pageref{lastpage}}
\maketitle

\begin{abstract}

Here we present the spatially resolved study of six Galactic planetary nebulae (PNe), namely IC~4593, Hen~2-186, Hen~2-429, NGC~3918, NGC~6543 and NGC~6905, from intermediate-resolution spectra of the 2.5~m Isaac Newton Telescope and the 1.54~m Danish telescope. The physical conditions (electron densities, N$_{e}$, and temperatures, T$_{e}$), chemical compositions and dominant excitation mechanisms for the different regions of these objects are derived, in an attempt to
go deeper on the knowledge of the low-ionization structures (LISs) hosted by these PNe. We reinforce the previous conclusions that LISs are characterized by lower (or at most equal) N$_{e}$ than their associated rims and shells. As for the T$_{e}$, we point out a \textit{possible} different trend between the N and O diagnostics. T$_e$[N~{\sc ii}] does not show significant variations throughout the nebular components, whereas T$_e$[O~{\sc iii}] appears to be slightly higher for LISs. The much larger uncertainties associated with the T$_e$[O~{\sc iii}] of LISs do not allow robust conclusions. Moreover, the chemical abundances show no variation from one to another PN components, not even contrasting LISs with rims and shells, as also found in a number of other works. By discussing the ionization photon flux due to shocks and stellar radiation, we explore the possible mechanisms responsible for the excitation of LISs. We argue that the presence of shocks in LISs is not negligible, although there is a strong dependence on the orientation of the host PNe and LISs.

\end{abstract}




\section{Introduction}

Planetary nebulae (PNe) represent the final stages on the evolution of low- and intermediate-mass stars. They are formed after the ejection of their outer envelopes, from the multiple stellar wind episodes occurred in the previous evolutionary stages and resulting in a complex bulk of ionized matter. Besides the 
large-scale components of PNe, such as shells, rims or halos, identified mainly from the emission of bright forbidden [O~{\sc iii}] together with H recombination lines, much smaller components are also recognized in PNe, due to their enhanced emission from low-ionization species, such as [N~{\sc ii}], [O~{\sc ii}], [S~{\sc ii}] or [O~{\sc i}] \citep[see e.g.][hereafter Paper~I]{1996A&A...313..913C,1998AJ....116..360B,2001ApJ...547..302G,2016MNRAS.455..930A}.  
These macro- and micro-structures are clearly recognizable in PNe' imaging catalogs such as \citet{1987AJ.....94..671B}, \citet{1992A&AS...96...23S}, \citet{1996iacm.book.....M}, \citet{1996A&A...313..913C} or \citet{1999A&AS..136..145G}.

\citet{2001ApJ...547..302G} compiled and classified these structures -- in terms of their morphology and kinematics -- as knots, filaments, jets and jet-like systems, which show up 
in axisymmetric pairs or isolated. To account for such variety of features, 
specific nomenclatures also appear in the literature \citep{1993ApJ...411..778B,1995ApJ...455L..63L,2000Ap&SS.274..205P}. 
Moreover, in the above quoted compilation the morphological and kinematic properties of low-ionization structures (LISs) of the 50 PNe sample were also compared with the predictions from theoretical models. The main conclusions then guided other studies with the aim of better constrain LISs' formation models. 
However, the enhanced low-ionization emission lines of LISs, relative to their surrounding medium, is still a perplexing issue, although observational evidence for their association with a not-insignificant molecular counterpart is becoming more and more appealing \citep{Reay1988}. For recent ideas on the relation between the intensity of low-ionization emission lines and the H$_2$ content of PNe, focusing the small-scale structures, see e.g. \citet[][]{2009MNRAS.398.2166G,2015ApJ...808..115M,Ramos_Larios2017,2017MNRAS.465.1289A,2020MNRAS.493.3800A}. 

Several observational works, sometimes with tailored models to explain the observations, have been conducted to 
characterise LISs' physical and chemical properties \citep[e.g.][]{1993ApJ...411..778B,1994ApJ...424..800B,1998AJ....116..360B,1997ApJ...487..304H,2003ApJ...597..975G,2004MNRAS.355...37G,2006MNRAS.365.1039G,2009MNRAS.398.2166G,2016MNRAS.455..930A,2016AJ....151...38D,2017PASA...34...36A,2020A&A...634A..47M,2021arXiv210505186M}.
The overall conclusions derived thus far are: (i) LISs occur in PNe of all different morphological types; (ii) LISs' electron temperatures are similar to those of the main nebular components; (iii) the electron densities of the main structures are higher than or equal to those measured for LISs; and (vi) there is no chemical abundance enrichment in LISs, i.e., all the nebular components have similar chemical composition. 

Even though, the LISs' kinematics were studied in detail for a number of PNe \citep[for instance in][]{2001ApJ...556..823S,2012ApJ...761..172G,Corradi1999,Corradi2000a,2000ApJ...542..861C,Akras2012,Derlopa2019}, a general knowledge about the kinematics of the LISs was missing until recently, when, at least for the case of pairs of jets/knots, this issue was solved. From a sample of 85 jets hosted by 58 PNe, \citet[]{Guerrero2020} found that jets can be divided into two populations: (i) those with spatial velocities below 100~km~s$^{-1}$, which represent $\sim$70\% of the sample, and (ii) those with mean velocities near 180~km~s$^{-1}$. Comparing the observed spatial and velocity distribution of jets, 
authors concluded that jets are mainly coeval with their parent PNe.

Regarding the knots, it is necessary to consider whether they occur in pairs or isolated, since it is counter-intuitive to necessarily link the formation of both kinds to the same  physical processes. For the latter, \citet{2009ApJ...700.1067M} convincingly showed that 
the isolated knots are part of the nebula's inner ring being swamp by the faster stellar wind. Similar conclusions were reached by \citet[][]{2015ApJ...808..115M} by studying the molecular waist of NGC~2346. These two results are not in conflict with the argument that isolated knots (or part of them) originate \textit{in situ} from the neutral AGB wind and are subsequently excited by shocks \citep[e.g.][]{2006ApJ...646L..61G} or by the central star radiation field \citep{1989MNRAS.241..625D,1998MNRAS.299..562S}. 

For the case of the pairs of knots, whilst a consensus has not yet been reached, the most promising recent models of magneto- or purely-hydrodynamic jet launching are converging to binaries with magnetic fields as the minimum requirement for the formation of collimated outflows. This is particularly true to account for highly collimated pairs of jets and knots. A number of works explore these arguments, and
we refer the readers to \citet[][]{2001ApJ...547..302G,2002ARA&A..40..439B, Guerrero2020} works, on which the theoretical efforts are reviewed, almost 2 decades apart. Moreover, the state of the art of such models (simulations) can be found in the following latest contributions: \citet[][]{2018ApJ...860...19G,2020ApJ...893..150G,2021ApJ...914..111G} as well as \citet[][]{2019ApJ...877...30B,2020ApJ...889...13B}.
Also to be mentioned is the series of studies starting with \citet[][]{2008MNRAS.391.1063A}, that simulate light-jets --~whose density are much lower than the density of the slow wind~-- which end up producing collimated PN shells and dense pairs of knots.

Explicitly meant to form jets/pairs of knots, \citet[]{2020ApJ...889...13B} had convincingly shown, via magneto-hydrodynamic simulations, that very dense axial knots formed in the slow, heavy flows that account for collimated pre-PNe and mature PNe eventually become the observed low-ionization knots, once their surfaces start to become ionized. This simulation  predicts that these LISs' high densities (10$^{5-7}$~cm$^{-3}$; their Fig.~6) and related UV opacities assure that LISs' interiors remain neutral and cold (3$\times$10$^{0-2}$~K; their Fig.~6); and with kinematics compatible with \citet[]{Guerrero2020} compilation. Such high densities in LISs have also been suggested for providing the necessary condition to self-shield the molecular hydrogen (H$_2$) recently discovered in some LISs from the stellar far-UV radiation \citep[][]{2013MNRAS.429..973M,2015ApJ...808..115M,2015MNRAS.452.2445F,2018ApJ...859...92F,2017MNRAS.465.1289A,2020MNRAS.493.3800A}.

In this paper, we present the analysis of spectroscopic data of a sample of 6 PNe with LISs, in order to obtain the spatial physico-chemical properties together with the excitation mechanisms present in these intriguing structures and their surrounding nebulae.
The observations and the procedure to analyse the data are described in section 2 and 3, respectively. In section~4 we present the results of spectroscopic analysis of the different structural components for the six PNe. Finally, in sections 5 and 6, we present the discussions and conclusions.

\section{Observations}

Low-resolution spectra of six PNe with embedded LISs were ob-
tained using the 2.5 m Isaac Newton Telescope (INT) at the Obser-
vatorio Roque de los Muchachos, Spain, and the 1.54 m Danish telescope at the European Southern
Observatory (ESO) at La Silla, Chile. INT and Danish data were taken, respectively, in August and September
of 2001, and April 1997.

The Intermediate Dispersion Spectrograph (IDS) mounted on the INT was used in conjunction with the 235~mm camera providing a spatial scale of 0.70~arcsec~pixel$^{-1}$ with the TEK5 CCD and the R300V grating, thus providing a spectral resolution of 3.3~\AA~pixel$^{-1}$ and a wavelength covering of 3650-7000~\AA. The slit width and length were 1.5~arcsec and 4~arcmin, respectively.

The Danish Faint Object Spectrograph mounted in the Danish telescope was used in conjunction with the DFOSC 2000$\times$2000 CDD camera, resulting in a spatial scale of 0.40~arcsec~pixel$^{-1}$, and the Grism$\#$4 (300 lines~mm$^{-1}$), which results in a wavelength range of 3600-8000~\AA~ and spectral resolution of 2.2~\AA~pixel$^{-1}$. The slit width was 1.0~arcsec and the slit length was $>$13.7~arcmin.

\begin{table}
\caption{Log of observations obtained with INT telescope in 2001 (1) and the Danish telescope in 1997 (2).} 
\label{tab:log}
{\small
\begin{tabular}{lcclcc}
\hline
PN name                           & Date       & PA   & Exposure {[}s{]}   & Seeing$^{\dag}$     &  Airmass        \\
\hline
\multirow{2}{*}{IC~4593$^{(1)}$}  & 08-29 & 62°  & 120, 300, 1200     &    $1.1$-$1.2$  &  1.21           \\
                                  & 08-29 & 139° & 120, 300, 1200     &    $1.0$-$1.2$  &  1.66           \\
\hline                          
Hen~2-186$^{(2)}$                 & 04-11 & 29°  & 60, 300, 1800      &    $2.1$        &  1.11           \\
\hline
Hen~2-429$^{(1)}$                 & 08-29 & 89°  & 1800               &    $1.8$        &  1.44           \\
\hline
\multirow{4}{*}{NGC~3918$^{(2)}$} & 04-11 & 30°  & 60,300             &    $1.6$        &  1.14           \\
                                  & 04-11 & 40°  & 300                &    $1.7$        &  1.17           \\
                                  & 04-10 & 42°  & 1800               &    $1.7$        &  1.18           \\
                                  & 04-10 & 70°  & 20, 300            &    $1.7$        &  1.39           \\
\hline                          
\multirow{2}{*}{NGC~6543$^{(1)}$} & 08-28 & 5°   & 20, 300            &    $1.3$-$1.4$    &  1.27           \\
                                  & 09-04 & 163° & 300, 1200          &    $1.2$-$1.3$    &  1.43           \\
\hline                          
NGC 6905$^{(1)}$                  & 08-31 & 161° & 300, 2700          &    $1.4$        &  1.17           \\
\hline
\end{tabular}
}
Note: $^{\dag}$Obtained through the FWHM of the stellar continuum measured in the spectra.
\end{table}

Several spectra have been obtained per PN, in different position angles (PA) and/or different exposure times (in order to avoid possible saturation of the usually brightest emission lines; e.g.  [O~{\sc iii}], H$\alpha$). The log of the observations is listed in Table~\ref{tab:log}. The reduction/analysis of the data was made using the {\sc longslit} package in {\sc IRAF} following the standard procedure: bias subtraction, flat-field correction and wavelength calibration using lamp frames. For the flux calibration of the spectra, spectro-photometric standard stars were observed and the flux calibration was also made with {\sc IRAF}. 

Observations were carried out taking into account the parallactic angle in order to avoid the differential chromatic refraction (DCR) effect, which mainly affects the blue side of the spectrum \citep[see e.g.,][]{2022MNRAS.512.4003M}. However, in some cases the PA used did not coincide with the parallactic angle and the DCR effect on these spectra was estimated. 
Taking into account the airmass (or sec(z)), altitudes of $\sim$2~km, and latitudes of $\sim\pm$30$^{\circ}$ per telescope, the DCR were derived from the equations of \citet[][]{1982PASP...94..715F}. Considering the bluest ([O~{\sc ii}]$\lambda3727$) and reddest ([S~{\sc ii}]$\lambda 6731$) lines of interest, we observe that the DCR effect could be present only in two cases. In IC 4593 (PA=139$^{\circ}$) and in NGC 3918 (PA=70$^{\circ}$), which amount at most to $\sim$1.9~arcsec and $\sim$1.4~arcsec, respectively. 
Nevertheless, comparing the results obtained for different PA and structures of these nebulae (see Tables~\ref{tab:ic4593_1} and \ref{tab:n3918_1}), we observe that DCR does not substantially affect the measurements on which we based our conclusions in this work.

\section{Nebular diagnostics}

First, the emission line fluxes were computed considering a Gaussian distribution in {\sc IRAF}. Then, the Nebular Empirical Analysis Tool \citep[NEAT,][]{2012MNRAS.422.3516W} was employed for the analysis. 
NEAT was used to identify the lines, compute the extinction coefficient (c$_{\beta}$), electron densities (N$_{e}$) and temperatures (T$_{e}$) as well as ionic (X$^{i+}$/H$^{+}$) and total abundances (X/H). NEAT uses the flux-weighted ratios of H$\alpha$, H$\gamma$ and H$\delta$ to H$\beta$ to calculate and correct the line fluxes for the interstellar extinction adopting the Galactic reddening law of \citet{1983MNRAS.203..301H}, with $R_{V}=A_{V}/E(B-V) = 3.1$.

The emission line intensities, 
in units of H$\beta$=100, together with c$_{\beta}$, N$_{e}$ and T$_{e}$ are listed in odd tables (\ref{tab:ic4593_1}-\ref{tab:n6905_1}) for several nebular components extracted from specific windows that are illustrated in Figures \ref{fig:ic4593} to \ref{fig:n6905}. Electron temperatures and densities were computed for different diagnostic lines. For the particular case of N$_{e}$ estimation using the [Ar~{\sc iv}] diagnostic lines, the theoretical value of He~{\sc i}~4713/4471=0.146 \citep{1999ApJ...514..307B} was used (considering T$_{e}$=10$^{4}$~K and N$_{e}$=10$^{4}$~cm$^{-3}$) to correct the blended [Ar~{\sc iv}]~$\lambda$4711+He~{\sc i}~$\lambda$4713 emission from the contribution of the He recombination line. 

For the ionic and total abundance calculations, as discussed by \citet{2012MNRAS.422.3516W}, the temperatures and densities that are more appropriate for the ionization potentials of the ionized gas are used. To correct for the unobserved ions, the ionization correction factors (ICF) of \citet{2014MNRAS.440..536D} were adopted, except for Ar/H and S/H, in which the ICFs from \citet[]{1994MNRAS.271..257K} were used. 
The ionic and total abundances, per PN nebular components, are listed in even tables (\ref{tab:ic4593_2}-\ref{tab:n6905_2}). For the Ne$^{2+}$/H$^{+}$ derived from the [Ne~{\sc iii}]~$\lambda$3967 line, we applied the correction for the contribution of the H$\epsilon$~$\lambda$3970 line, considering the theoretical ratio H$\epsilon$/H$\beta \sim 0.158$ 
\citep{2006agna.book.....O}.

It should be noted that NEAT uses a Monte Carlo scheme to calculate the statistical uncertainties of the parameters, based on the flux uncertainties and their propagation through the diagnostics~\footnote{For further information about this tool and its error propagation, readers are directed to  
\citet{2012MNRAS.422.3516W}.}. 
Errors related with N$_{e}$ and T$_{e}$ were obtained directly from the uncertainties of the lines involved on the diagnostic ratio, as in the previous papers of the series \citep[][]{2003ApJ...597..975G,2004MNRAS.355...37G,2009MNRAS.398.2166G,2016MNRAS.455..930A,2020Galax...8...46M}.

\section{Results}

In the next subsections we present the 
spectroscopic analysis of six PNe, for the different morphological structures -- LISs, shells, rims and a portion of the nebula along the various slit positions, generally of high-ionization, labeled as Neb. We also summarize the main results 
from the literature.

Figures \ref{fig:ic4593} to \ref{fig:n3918}, \ref{fig:n6543} and \ref{fig:n6905} show the above defined components, per nebula. 
The normalized flux distribution of the [N~{\sc ii}]$\lambda$~6584 (solid-line) and [O~{\sc iii}]$\lambda$~5007 (dashed-line) 
along the slits, as well as the ratio of these emission lines, are presented in the middle and lower panels of the figures. The line-ratio plots are extremely useful to define the the regions of the spectra to extract the regions under analysis. Ideally, the underlying nebular emission should be subtracted from the LISs emission lines, to insure that LISs' properties are not contaminated by those of the large-scale nebular structures. However, with the current data, this is a cumbersome task. This kind of correction can be better addressed with IFU data, by computing the large-scale nebula structures emission from the surrounding region. A good example of such correction is by \citet[][]{2022MNRAS.512.4003M} who managed to resolve the inner region from the surrounding nebular emission in the velocity space. Therefore, we compute the emission lines from different nebular structures such as Neb, LISs and rims/shells, and compare the outcomes among these structures.

For the six PNe, the emission line fluxes, absolute flux of H$\beta$ (observed), c$_{\beta}$, N$_{e}$ and T$_{e}$ 
are presented in the odd tables (from Table~\ref{tab:ic4593_1} to \ref{tab:n6905_1}). On the other hand, the 
ionic and total abundances of the nebular components 
are show on the even tables (from Table~\ref{tab:ic4593_2} to \ref{tab:n6905_2}). The literature with which we compared the results of our diagnostics are listed in the captions of  Figures \ref{fig:abundances1} and \ref{fig:abundances2}.

\subsection{IC 4593}

\begin{figure} 
    \centering
	\fbox{\includegraphics[width=0.7\columnwidth]{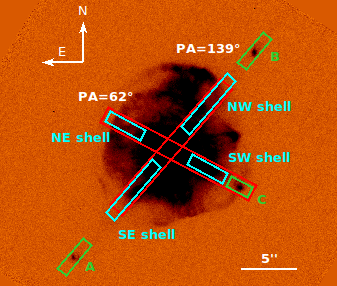}}
	\includegraphics[width=0.75\columnwidth]{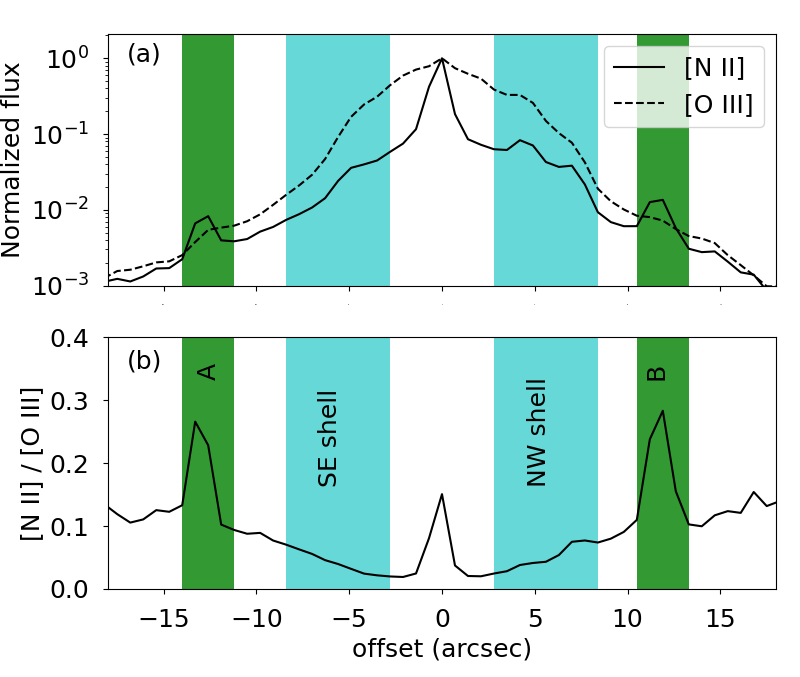}
	\includegraphics[width=0.75\columnwidth]{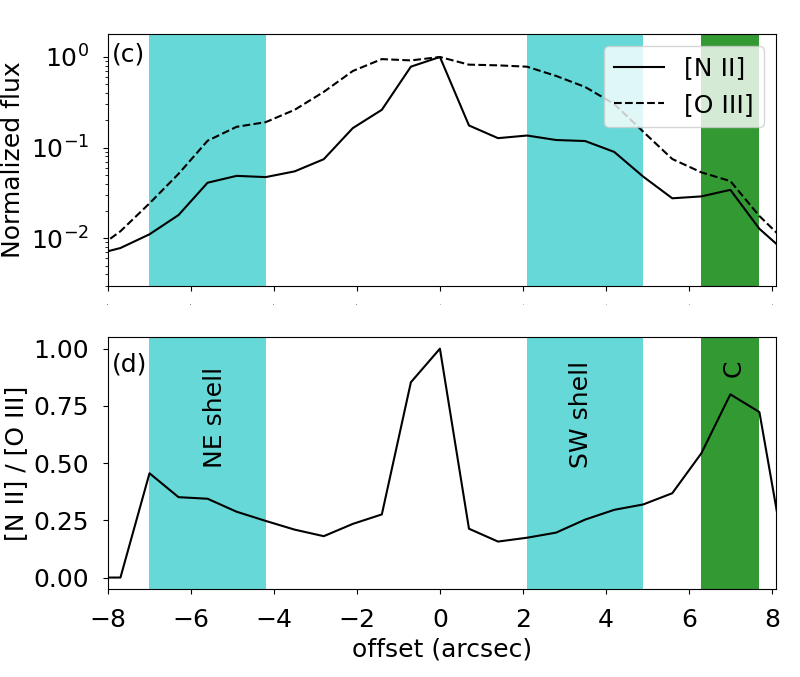}
        \caption{\textit{Upper panel}: HST [N {\sc ii}] image of IC~4593. All regions and nebular components selected for the spectroscopic analysis, LISs, shells and Neb are highlighted with green, cyan and red coloured boxes, respectively. The extracted window of the LISs is 3.5~arcsec for the PA=139$^{\circ}$ slits (regions A and B) and 2.1 for the PA=62$^{\circ}$ slit (region C), for the shells is 3.5~arcsec for the slit at PA=62$^{\circ}$ and 6.3~arcsec for the slit at PA=139$^{\circ}$ and 15~arcsec for both Neb regions regardless the slit position. The size of the image is 30$\times$25 arcsec$^{2}$. \textit{Lower panels}: the flux distribution of the [N{\sc ii}] and [O{\sc iii}] emission lines, normalized to 1.0 -- in logarithmic scale -- and the [N {\sc ii}]/[O {\sc iii}] line ratio, along the slits at PA=139$^{\circ}$ (panels a and b) and PA=62$^{\circ}$ (panels c and d). The coloured regions correspond to the equivalent region/component in the upper panel.}
    
    \label{fig:ic4593}
\end{figure}

In the optical, IC 4593 is a complex multiple-shell PN, with a roundish bright filamentary outer shell, with $\sim$16'' in extension. The (H$\alpha$+{[N \sc{ii}]})/{[O \sc{iii}]} ratio images from \citet{1996A&A...313..913C} revealed the presence of a pair of knots, which authors called jet-like features, located near $\sim$12'' from the center and oriented along PA=139$^{\circ}$, as well as an isolated knot embedded in the inner layers along PA=62$^{\circ}$ (Fig.~\ref{fig:ic4593}). According to \citet{Corradi1997} the jet-like features have radial velocities of 2~km~s$^{-1}$, whereas for the isolated knot this value is of 1~km~s$^{-1}$. 

For the analysis of this nebula we selected the slit which contains the pair of LISs (PA=139$^{\circ}$) and the isolated one (PA=62$^{\circ}$), together with the inner shell (4 regions) and Neb. We, previously, studied this nebula and the results were presented in \citet[][]{2020Galax...8...46M}, however, the emission-line fluxes were remeasured in an attempt to reduce uncertainties. The improved results are presented here.

The interstellar extinction coefficient of IC~4593 varies significantly between the components and slit positions, from 0.02 to 0.22, with an average value of $0.11\pm0.02$. Interestingly, a similar wide range of c$_{\beta}$ is found in the literature, from  0.01 \citep[along E-W direction]{1996A&AS..116..249C}, 0.05 \citep[]{1992A&AS...95..337T}, 0.125 \citep[]{1978ApJ...219..914B}, 0.17 \citep[along E-W direction]{2005ApJS..157..371R} up to 0.24$\pm$0.16 \citep[PA=0]{2009ApJ...694.1335D}.
This wide range of c$_{\beta}$ values likely indicates a significant variation across the nebula, which could, for instance, be associated with mass loss variations or
the dust ejected in the AGB phase \citep[e.g.][]{2016A&A...588A.106W}.
Integral field spectroscopy should be very helpful to investigate the variation of c$_{\beta}$ in both spatial directions of this nebula.

From the IC~4593's spectra it was possible to estimate the N$_{e}$ and T$_{e}$ using the diagnostic line ratios of sulphur, nitrogen and oxygen (see Table~\ref{tab:ic4593_1}). Considering both slit positions, T$_{e}${[O~\sc{iii}]} varies from (8100$\pm$1800)~K to (8800$\pm$1100)~K. Taking into account the uncertainties of T$_{e}$, our results are in agreement with the previously published values  (for the literature references, see caption of Fig.~\ref{fig:abundances1}). The only exception is the pair of LISs along 139$^{\circ}$, which has higher values exceeding 11000~K. However, due to the large relative errors, we cannot argue for any variation in T$_{e}${[O~\sc{iii}]} between the different components. In the case of T$_{e}${[N~\sc{ii}]}, the values computed in this work vary between 9510~K and 11400~K, again in agreement with the literature. Here as well, the LISs at PA 139$^{\circ}$ are characterized by higher values ($\sim$13000~K) and higher uncertainties. On the other hand, N$_{e}$ is solely determined from the {[S~\sc{ii}]} doublet lines and takes values from 2100$\pm$1200~cm$^{-3}$ to 2800$\pm$600~cm$^{-3}$ that agrees with some published values, but is higher than others, as shown in Fig.~\ref{fig:abundances1}. According to the analysis above, we can see that both T$_{e}$ and N$_{e}$ do not show any significant spatial variation among either the shells and LISs, or even from one to another PA (Neb).

The ionic and total abundances for IC~4593 are listed in Table~\ref{tab:ic4593_2}. 
No variation in He abundance between components is observed. The average value is 0.100$\pm$0.004, in good agreement with the majority of the literature (see Fig.~\ref{fig:abundances1}), but lower than 0.11 reported by \citep{1998ApJ...493..247K} or 0.127 by \citet{2006ApJ...651..898S}. As for O/H, there is also no trend among nebular components, except for both LISs located along 139$^{\circ}$ that show lower values, variation related to the face values higher T$_{e}$~[O~{\sc iii}] of these LISs{\footnote{We note that, at variance with T$_{e}$ and N$_{e}$ uncertainties, obtained directly from the propagation of the emission-line ones, the ionic and total abundance uncertainties were estimated via Monte Carlo approach, within NEAT. This is why X/H uncertainties seems to be small, even when the T$_e$ uncertainties are severe.}}.
Comparing with the published abundances, we note that our results of O/H are sometimes higher but, considering the errors, in agreement with \citet{1996RMxAA..32...47B} and \citet{2006ApJ...651..898S}. For N/H our results are similar to those reported by \citet[]{1996RMxAA..32...47B} and \citet[]{1998ApJ...493..247K}, and higher than those achieved in other works. 
Our results could indicate a possible variation in N/H abundances through the structures. However,  \citet{1998ApJ...501..221G} while analyzing the empirical 
abundances determinations, argued that abundances estimations through line intensities depend on the line of sight. 
Hence, the empirical abundances at different regions in a nebula may not be representative. In fact, we obtain a variation in the N/O abundance ratio, with the highest values found for the LISs along 139$^{\circ}$. This result is, though, closely correlated with the problem of the much higher face values of the [O~{\sc iii}] temperature, as discussed above.
Ne/H abundances derived in this work are higher than in the literature. On the other hand, when performing the Ne/O ratio, the values fall close to those published by \citet{1978ApJ...220..193B,1996RMxAA..32...47B,2006ApJ...651..898S}, within 
errors. 
Our Ar/H abundance is lower than the literature, while Cl/H abundance has not yet been reported. Following the criteria giving by \citet{1978IAUS...76..215P} and \citet{1994MNRAS.271..257K} to define Type I PNe -- as He/H$\geq$0.125 and N/O$\geq$0.5 -- this nebula is classified as non-Type I PN.

\subsection{Hen 2-186}

\begin{figure} 
    \centering
    \fbox{\includegraphics[width=0.7\columnwidth]{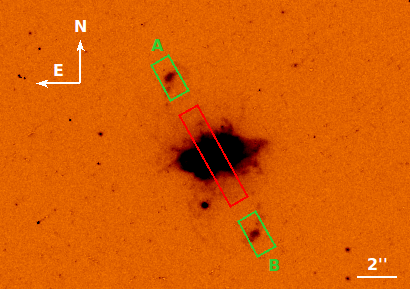}}
    \includegraphics[width=0.75\columnwidth]{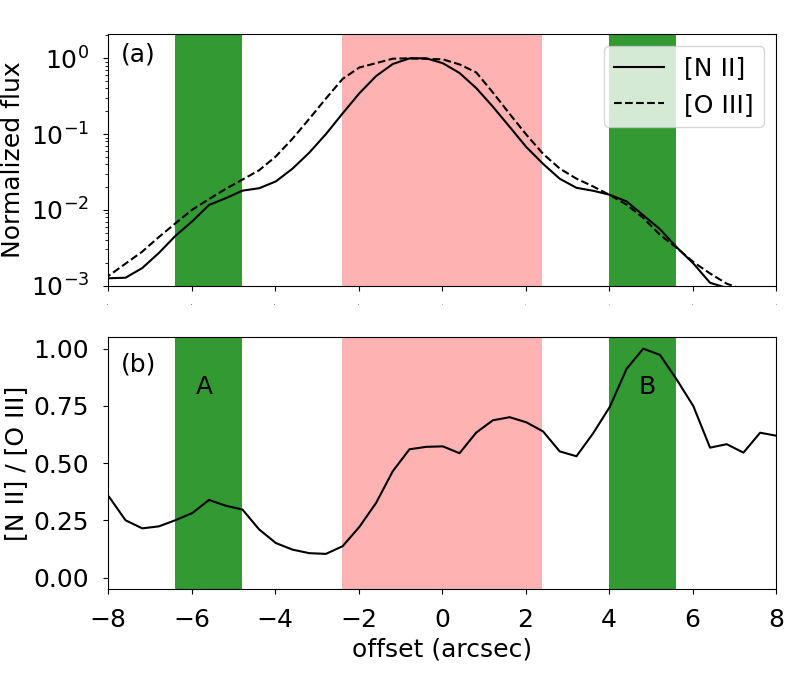}
    \caption{\textit{Upper panel:} HST [N {\sc ii}] image of Hen 2-186. The size of the field is 20$\times$14 arcsec$^{2}$. The nebular components under analysis are indicated by the boxes: the nebular region (Neb, 5.2 arcsec, in red) and the LISs (A and B, 2 arcsec, in green). \textit{Lower panels:} Same as Fig~\ref{fig:ic4593}.}
    \label{fig:hen2-186}
\end{figure}

Hen 2-186 is a relatively small
PN with an angular extension of $\sim$3.5''. The (H$\alpha$+{[N \sc{ii}]})/{[O \sc{iii}]} ratio images from \citet{1996A&A...313..913C} have revealed a pair of LISs about $\sim$4.5'' from the center, and oriented along 29$^{\circ}$ in position angle. The clump located near to the center of the nebula is actually a field star. Three components were selected for the analysis of this nebula: the pair of LISs and the inner nebula (Neb), all of them along the PA of 29$^{\circ}$ (see Fig.~\ref{fig:hen2-186}). This particular nebula belongs to a specific group of PNe, whose jets (LISs) have radial velocities greater than 100~km~s$^{-1}$,
with a value of 135~km~s$^{-1}$ \citep{Corradi2000a,Guerrero2020}. \citet{Corradi2000a} suggested these LISs 
might be prominent point-symmetrical features within a more general structure containing faint bipolar lobes.

Table~\ref{tab:h186_1} shows the diagnostics derived for the different  structures of the nebula. c$_\beta$ of Hen~2-186 has an average value of 0.70$\pm$0.04, which agrees with the most recent estimation from \citet[c$_{\beta}$=0.75]{2010RMxAA..46..159C}, but is higher than that presented by \citet[c$_{\beta}=0.4$]{1970ApJ...160..887K} and lower than the value of 0.93 from \citet{1992A&AS...95..337T}. 
The Neb component has a T$_{e}$[O~{\sc iii}] of 14600$\pm$330~K, slightly higher than T$_{e}$[N~{\sc ii}]=11300$\pm$240~K. These figures coincide, within the uncertainties, with those presented so far in the literature (see Fig.~\ref{fig:abundances1}). We were also able to derive T$_{e}$[O~{\sc i}], though with larger uncertainty, as being 12700$\pm$3200~K. As for the N$_{e}$, we estimated N$_{e}$[Cl~{\sc iii}] of 8100$\pm$1600~cm$^{-3}$, N$_{e}$[Ar~{\sc iv}] of 5420$\pm$210~cm$^{-3}$ and N$_{e}$[S~{\sc ii}] of 3990$\pm$130~cm$^{-3}$, for Neb. The latter is in good agreement with that reported by \citet{2010RMxAA..46..159C}. Taking into account the 
LISs and Neb T$_{e}$, we argue that Hen~2-186 does not reveal any important temperature spatial variation. With regard to the N$_{e}$, LISs are found to be less dense than Neb, by a factor between $\sim$2.3 and $\sim$3.7. The higher densities found using the argon and chlorine line ratios may imply a density stratification in the inner nebula.  Further spatially-resolved analysis is needed to confirm this trend.

Hen~2-186's ionic and total abundances are presented on Table~\ref{tab:h186_2}. He abundances are found  unchanged among the nebular components,  with an average of 0.133$\pm$0.009, in good agreement with the literature (see Fig.~\ref{fig:abundances1}). For the N abundance no variation is detected through the components, and the Neb result coincides with that published by \citet{2010RMxAA..46..159C}, but is lower than that reported by \citet{2017MNRAS.471.4648V}. Nonetheless, for O/H we can see that the LISs exhibit higher values than the nebula by a factor between $\sim$2.1 and $\sim$2.5 \footnote{An over- or under-abundance 
of around two is not high enough to allow a firm conclusion of abundance variation across the 
nebula, due to the cavities of the ICF scheme \citep{1994MNRAS.271..257K,2014MNRAS.440..536D}.}, and if errors are taken into account Neb value matches the published ones (see the caption of Fig.~\ref{fig:abundances1}). 
A similar behavior is observed for Ne/H, for which LISs show higher abundances by a factor between $\sim$1.9 and $\sim$2.3. With regard to Ar/H, it can be seen that it does not vary significantly among regions, and our estimations are lower than those reported in literature. On the other hand, the average of S/H abundance is similar to previous works. According to its chemical abundances, Hen 2-186 can be classify as non-type I PNe.

\subsection{Hen 2-429}

\begin{figure}
    \centering
    \fbox{\includegraphics[width=0.7\columnwidth]{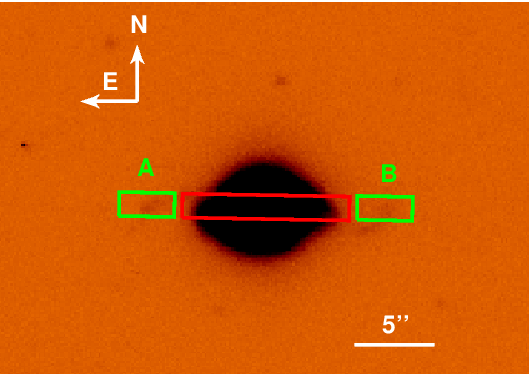}}
    \includegraphics[width=0.75\columnwidth]{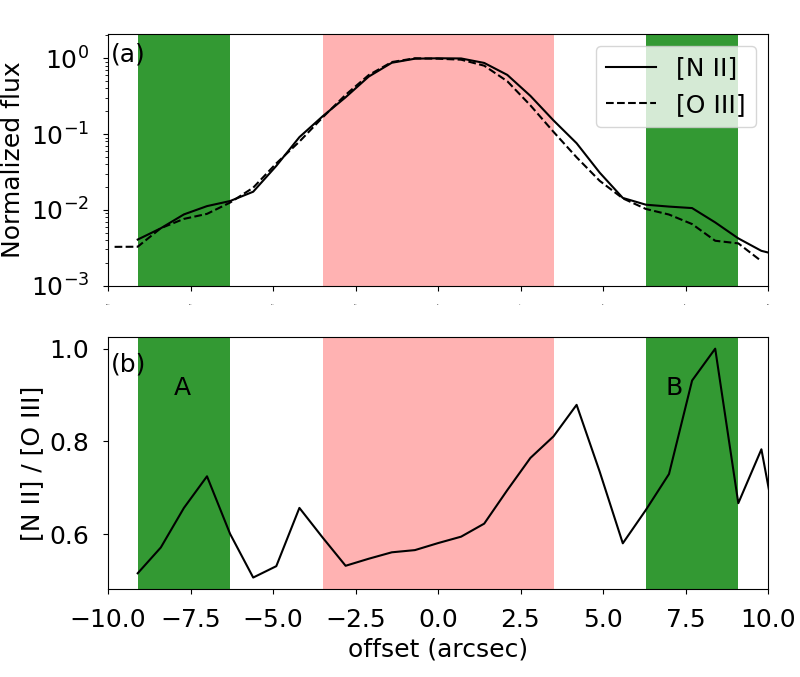}
    \caption{\textit{Upper panel:} [N {\sc ii}] image of Hen 2-429 from \citet{1996iacm.book.....M}. The size of the field is 33$\times$23 arcsec$^{2}$. The nebular components under analysis are indicated by the boxes: the nebular region (Neb, 10.5 arcsec, in red) and the LISs (A and B, 3.5 arcsec, in green). \textit{Lower panels:} Same as previous Figures. 
    Note: the bottom panel, which shows the line ratio, shows peaks that are not related to the features of the nebula, but to the displacement between spatial distribution of the emission lines. The  [N{\sc ii}] and [O{\sc iii}] spectra are  misaligned for a fraction of pixel, which generates the false peaks  near the continuum, within the Neb region.}
    \label{fig:hen2-429}

\end{figure}

The PN Hen 2-429 possesses a point-symmetric morphology with an elliptical shell and a pair of faint jet-like features, the latter only prominent in low-ionization emission \citep[]{1999AJ....117..967G}. These LISs are located at $\sim$6.5'' from the center, oriented along the PA of 89$^{\circ}$, and have systemic radial velocities of 5~km~s$^{-1}$, according to \citet{Guerrero2020}. Our study of this nebula selects three components: the pair of LISs and the main body of the nebula (Neb), all of them along  the position angle of 89$^{\circ}$, as it can be seen in  Fig.~\ref{fig:hen2-429}.

Our results in terms of the spectroscopic characterization (line intensities,  H$\beta$ flux, c$_{\beta}$, N$_{e}$ and T$_{e}$) of He~2-429's components are given in Table~\ref{tab:h429_1}. As the table shows, its extinction coefficient is the highest in our sample, with an average value of 2.08$\pm$0.09, 
which agrees with (or is close enough to) the values reported in \citet[2.12]{2011A&A...525A..58G}, \citet[2.3]{1992A&AS...95..337T} and \citet[2.21]{2007A&A...463..265G}.  
The electron temperatures of Hen~2-429 are provided only for Neb, T$_{e}$[O {\sc iii}] of 9800$\pm$2400~K and T$_{e}$[N {\sc ii}] of 9400$\pm$700~K, as neither the [N {\sc ii}] $\lambda$5755 nor the [O {\sc iii}] $\lambda$4363 auroral lines were detected in the LISs. Our estimations are consistent, within the errors, with the previous ones reported in the literature (see Fig.\ref{fig:abundances1}). As for N$_{e}$, we derived N$_{e}$[S~{\sc ii}] of 5710$\pm$270~cm$^{-3}$ and N$_{e}$[Cl {\sc iii}] of 5200$\pm$2400~cm$^{-3}$ for the Neb component, values that are slightly smaller than
previously reported. Compared to Neb, the N$_{e}$ of the LISs are lower by a factor of $\sim$1.8 and $\sim$1.3 for A and B, respectively.

Ionic and total abundances are shown in Table~\ref{tab:h429_2}. Taking into account the errors, Neb abundances are in good agreement with \citet{2007A&A...463..265G}, with the exception of He/H for which our value (0.151$\pm$0.019) is higher than the published one (see Fig.~\ref{fig:abundances1}). As it can be seen in Table~\ref{tab:h429_1}, no electron temperature was 
computed for the LISs. However, the abundances of LISs were estimated adopting the mean- and low-ionization temperatures of the Neb. Within the uncertainties, we find no significant variation of abundances between LISs and Neb. Values on Table~\ref{tab:h429_2} allow us to classify Hen 2-429 as non-type I PN.

\subsection{NGC 3918}

\begin{figure}
    \centering
    \fbox{\includegraphics[width=0.7\columnwidth]{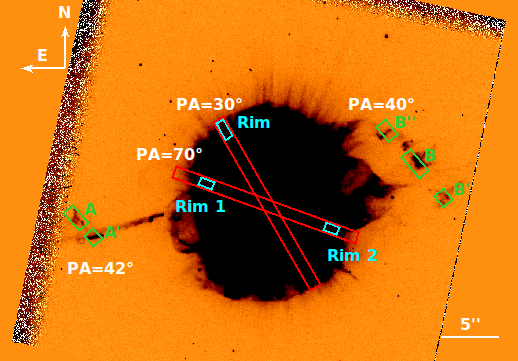}}
    \caption{HST [N {\sc ii}] image of NGC 3918. The size of the field is 44$\times$31 arcsec$^{2}$. The nebular components under analysis are indicated by the boxes: the nebular regions (Neb, same extension in all directions: 16.8 arcsec, in red); the rims (in cyan; for PA=30$^{\circ}$ 1.6 arcsec and for PA=70$^{\circ}$ 1.2 arcsec); the LISs (in green; for A 2 arcsec, for A' 1.2 arcsec, for B 2.4 arcsec, for B' 1.2 arcsec and for B'' 1.6 arcsec).}
    \label{fig:n3918}
\end{figure}

NGC 3918 is a widely studied planetary nebula and characterized by a complex morphology -- various models have been used to reproduce its morphological properties. \citet{1987ApJ...314..551C} proposed a biconical geometry, whereas \citet{2003MNRAS.340.1153E} adopted, besides the biconical one, two spindle-like models. Both representations have integrated emission-line spectra which are in agreement with the observations. \citet{2017MNRAS.472.1182P} came to the conclusion that NGC 3918 has a complex point-symmetric morphology. 
Aside from the large-scale structures, it has LISs located outside the main PN body and oriented approximately along the major axis. (H$\alpha$+[N~{\sc ii}])/[O~{\sc iii}] ratio images from \citet{1996A&A...313..913C} initially revealed the presence of one jet and one knot on radially opposite sides. In this work, we present the spectroscopic analysis of several NGC 3918's structures, by adopting the nomenclature defined by \citet{Corradi1999}, including also two other micro-structures named B' and B''. All nebular components under our analysis are shown in the HST [N~{\sc ii}] image of the nebula in Fig.~\ref{fig:n3918}. 

\begin{figure}
    \centering
    \includegraphics[width=0.49\columnwidth]{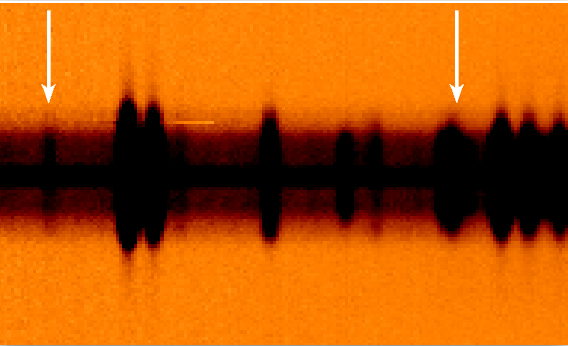}
    \includegraphics[width=0.49\columnwidth]{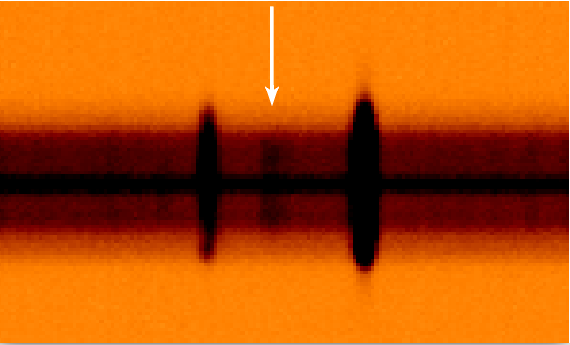}
    \caption{Two different regions of the 2D spectra of NGC 3918 to highlight the mimics of emission-line stars. \textit{Left panel:} the arrows correspond to the lines of C~{\sc ii} at $\lambda$4267~\AA~ (left) and C~{\sc iii} at $\lambda$4647~\AA~ (right). \textit{Right panel:} the arrow indicates the position of the C~{\sc iv} at $\lambda$5801/12~\AA~ line.}
    \label{fig:mimics}
\end{figure}

The properties that the present data allow us to derive, for the different nebular components, are shown in Table~\ref{tab:n3918_1}. Note that the recombination lines of C~{\sc ii} at $\lambda$4267~\AA~, C~{\sc iii} at $\lambda$4647~\AA~ and C~{\sc iv} at $\lambda$5801/12~\AA~ are detected in our spectra and also reported by \citet{1987ApJ...314..551C} and \citet{2014A&A...570A..26G}. In the latter, these recombination lines were considered as \textit{mimics of emission-line stars}. As \citet[]{2014A&A...570A..26G}, we also suggest that the origin of these lines is not stellar, 
and base our argument on their spatial distribution in the 2D spectra (see Fig.~\ref{fig:mimics}). 
 
The interstellar extinction 
of NGC~3918 has an average value of 0.28$\pm$0.01, without a significant variation among the different structures. 
Our value is lower than those obtained by \citet{1985RMxAA..11...35P}, \citet{2003MNRAS.345..186T}, and \citet{2005ApJS..157..371R}, which amount to 0.40, 0.44 and 0.40, respectively. Nevertheless, our result is consistent, within the uncertainties, with the values found in \citet[0.33$\pm$0.14]{1987ApJ...314..551C} (using the Balmer decrement), \citet[0.27]{2003PASP..115...80K} and \citet[0.26$\pm$0.06]{2015MNRAS.452.2606G}. N$_{e}$ and T$_{e}$ were estimated using the diagnostic line ratios of sulphur, chlorine, argon, nitrogen and oxygen. No significant variation is found for T$_{e}$[N~{\sc ii}] among the different structures. For Neb we obtained T$_{e}$[N~{\sc ii}] of 11100$\pm$500~K, along 30$^{\circ}$, and 11200$\pm$800~K for the PA of 70$^{\circ}$, both agree well with the literature. The average value of T$_{e}$[O~{\sc iii}] for Neb is 12700$\pm$140~K, which is also in good agreement with the published values. This latter temperature also has no significant variation across the nebular components, though LIS B'', along the PA of 40$^{\circ}$, has higher T$_{e}$[O~{\sc iii}] face values, with far higher errors than the other structures. As for the N$_{e}$, we computed 5230$\pm$190~cm$^{-3}$, 5700$\pm$800~cm$^{-3}$ and 4220$\pm$80~cm$^{-3}$ as the average values for Neb, from the [S~{\sc ii}], [Cl~{\sc iii}] and [Ar~{\sc iv}] diagnostic lines, respectively.  N$_{e}$[S~{\sc ii}] is consistent with the literature, but lower than the value determined by \citet[5700~cm$^{-3}$]{1991ApJ...377..210R} and higher than that of \citet[3800$\pm$380~cm$^{-3}$]{2003PASP..115...80K}. The mean value of 
N$_{e}$[Cl~{\sc iii}] is also in good agreement with the literature, whereas for the case of N$_{e}$[Ar~{\sc iv}] our value is
lower than those published by \citet{1989ApJ...343..811S} and \citet{,2015MNRAS.452.2606G},  of 6166~cm$^{-3}$ and 6500$^{+1300}_{-1200}$~cm$^{-3}$, respectively. Regarding the variation among different nebular components, it should be noted that for the case of LISs, the N$_{e}$ could only be calculated using the sulfur diagnostic lines that, in comparison with Neb, is lower by a factor that varies from $\sim$2.6 to $\sim$3.9.

Ionic and total abundances are listed in Table~\ref{tab:n3918_2}, from which no variation in He abundances is observed. 
The average He/H is 0.104$\pm$0.003, in good agreement with the literature (see Fig.~\ref{fig:abundances2}). For O/H there is no much contrast among nebular components, except for the LIS B'', along 40$^{\circ}$, for which we find a value of approximately half that of the other structures. Similarly to the case of IC~4593, this under-abundance of B'' is related with the much higher, and much uncertain, T$_{e}$[O~{\sc iii}]. 
Excluding this LIS, the average O/H is (4.0$\pm$0.1)$\times 10^{-4}$, which is lower than those obtained previously by some authors, 
but considering the errors agrees with \citet{2018MNRAS.473..241H}. The N/O abundance ratio also agrees with the published values in the literature. Considering the quantities in Table~\ref{tab:n3918_2}, we conclude that this nebula classifies as non-type I. The neon abundance behaves as O/H, without any variation among nebular components (with the exception of  B'', whose value is lower than the average of the other structures, by a factor $\sim$5.7). Given that B'' is the faintest nebular component of NGC~3918, and 
also the fact the its behaviour differs significantly from that of the other LISs of the nebula, we do not argue that the variation in either T$_{e}$ or X/H is real. 
At variance with Ar abundance, the average Cl/H and S/H are consistent with those published by \citet{2003PASP..115...80K}. 

\subsection{NGC 6543}

NGC~6543 is among the most widely studied PNe and several works have focused on understanding its complex morphology \citep[e.g.][]{1987AJ.....94..948B,Miranda1992,1994ApJ...424..800B,Reed1999,2016MNRAS.462..610R,2020MNRAS.495.2234G} concluding that it is formed by a geometrically thick expanding ellipsoid with two bright shells. NGC~6543 also posses a pair of LISs with radial velocities of 39~km~s$^{-1}$ \citep{Miranda1992,Reed1999,Guerrero2020}. Moreover, recent near-IR narrow-band imagery of the nebula has revealed the presence of H$_2$ emission in the [N~{\sc ii}] emission between the rims/shells and LISs \citep[][]{2020MNRAS.493.3800A}.

\begin{figure}
    \centering
    \fbox{\includegraphics[width=0.7\columnwidth]{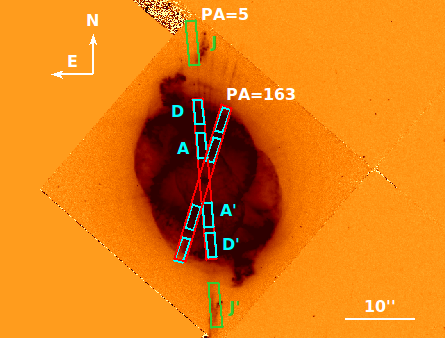}}
    \includegraphics[width=0.75\columnwidth]{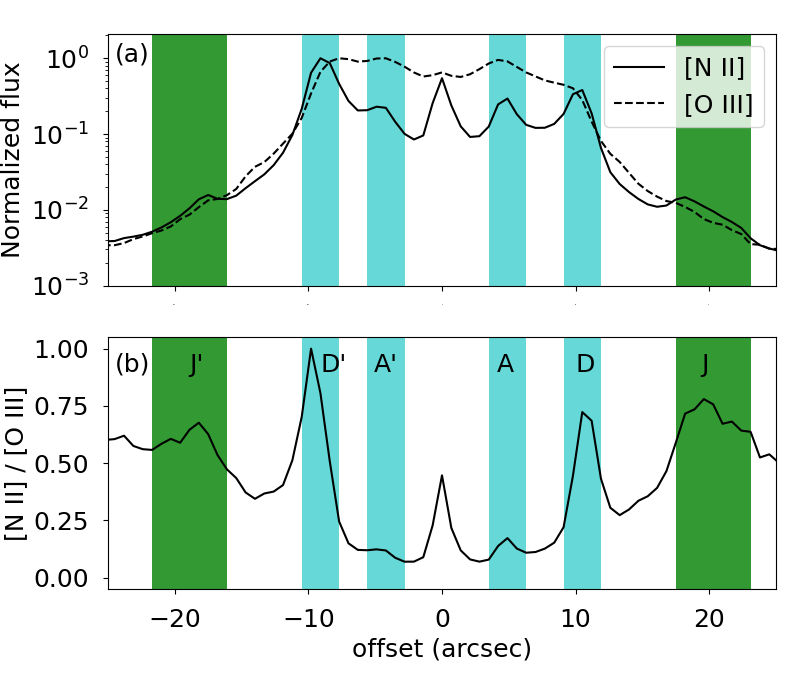}
    \caption{\textit{Upper panel:} HST [N {\sc ii}] image of NGC 6543. The size of the field is 63$\times$48 arcsec$^{2}$. The nebular components under analysis are indicated by the boxes: the nebular regions (Neb, same extension in both directions: 23.1 arcsec, in red); the shells (A, A', D, D', 3.5 arcsec in cyan); the LISs (J and J', 6.3 arcsec, in green). \textit{Lower panels:} Same as previous Figures, along the slit which contains the LISs, PA=5$^{\circ}$.}
    \label{fig:n6543}
\end{figure}

The analysis we present here follows the same nomenclature proposed by \citet{Miranda1992}. Our study is based on two slit positions (5$^{\circ}$ and 163$^{\circ}$), and the nebular components under analysis are denoted as A, D and J (in the northern half), with their counterparts A', D', J', and Neb. The J and J' LISs are covered by the slit at PA of 5$^{\circ}$ (see Fig.~\ref{fig:n6543}).

Table~\ref{tab:n6543_1} shows the line intensities, observed H$\beta$ flux, c$_{\beta}$, N$_{e}$ and T$_{e}$ derived for the above structures, along both slits. 
From Table~\ref{tab:n6543_1} it is straightforward noticing that c$_{\beta}$ 
varies among the components, from $\sim$0.02 up to 0.26. The average extinction coefficient is 0.11$\pm$0.02, which is in good agreement with  \citet[0.12]{1999A&A...347..967P} and \citet[0.14]{2008ApJ...677.1100W}. Interestingly, c$_{\beta}$ values in the range from 0.08 up to 0.3 have also been reported by other authors as \citet[][0.08]{2005ApJS..157..371R}, \citet[][0.1]{2004MNRAS.351.1026W}, \citet[][0.22]{1970ApJ...160..887K} and \citet[][0.3]{2000MNRAS.318...77H}. As in the case of IC~4593, this non-negligible variation in c$_{\beta}$ points to spatial variations within the nebula. Such a spatial variation is also obtained, for instance, from the 2D c$_{\beta}$ map of NGC~7009 computed for MUSE IFU data \citep{Walsh2018}.

For the estimation of electron temperature and density, the diagnostic line ratios of sulphur, chlorine, argon, nitrogen and oxygen were used. Considering all nebular components, except the LISs, along both slit positions, the computed average value of T$_{e}$[O~{\sc iii}] is 8150$\pm$210~K 
in very good agreement with the literature (see Fig.~\ref{fig:abundances2}). As for the J and J' (LISs), T$_{e}$[O~{\sc iii}] is 12100$\pm$7500~K and 10000$\pm$7000~K, respectively. These results may be indicative of extra heating mechanisms, such as shocks. However, the much higher uncertainties obtained for the LISs' do not allow firm conclusions about the possibility of real variations of  T$_{e}$[O~{\sc iii}] among the nebular components. Regarding the T$_{e}$[N~{\sc ii}] of components A, A', D and D', we obtain an average value of 10000$\pm$500~K, in good agreement with previous estimations. As for the J and J' contrast with the other components, the errors again prevent robust conclusions. \cite{1994ApJ...424..800B} also studied three different regions of this nebula, named rim, cap and ansae (the latter are equivalent 
to our LISs), with a long-slit along the position angle of 14$^{\circ}$. \cite{1994ApJ...424..800B}  electron temperatures for the three structures are T$_{e}$[O~{\sc iii}] (T$_{e}$[N~{\sc ii}]) of 8000~K (9300~K), 7900~K (9000~K) and 8200~K (7400~K), respectively. Therefore, for the rim and cap components their values are coincident with ours, while for the ansae our results are higher, with higher uncertainties, though strictly speaking, ours and \cite{1994ApJ...424..800B} derived quantities are in good agreement.

Sulfur, chlorine and argon lines allowed the derivation of the electron densities. A subtle variation is found among the nebular components, but J/J', the LISs, are showing much lower electron densities. In particular, the average value for N$_{e}$[S~{\sc ii}] is 5060$\pm$100~cm$^{-3}$ for the slit at along 5$^{\circ}$ and 5220$\pm$130~cm$^{-3}$ for the 163$^{\circ}$ one. As for the J and J' LISs, we determined N$_{e}$[S~{\sc ii}] of 950$\pm$240~cm$^{-3}$ and 930$\pm$210~cm$^{-3}$, respectively. These values are lower when compared to the other components, by factors of $\sim$5.3 for J and $\sim$5.5 for J'. The comparison with the literature shows that our densities are close to the previous calculations (see Fig.~\ref{fig:abundances2}). Moreover, \citet{1994ApJ...424..800B} computed the electron density for the rim, cap and ansae obtaining 4600~cm$^{-3}$, 5000~cm$^{-3}$ and 2200~cm$^{-3}$, respectively. N$_{e}$[Cl~{\sc iii}] and N$_{e}$[Ar~{\sc iv}] were also computed for the main nebular structures (Neb) and their average are 4630$\pm$310~cm$^{-3}$ and 6800$\pm$900~cm$^{-3}$, respectively. The resultant values again are in agreement with the previous estimations of N$_{e}$[Cl~{\sc iii}] of  \citet[4660~cm$^{-3}$]{2004MNRAS.351.1026W} and \citet[5000$^{+2100}_{-1400}$~cm$^{-3}$]{2008ApJ...677.1100W}, but slightly larger than N$_{e}$[Ar~{\sc iv}] of \citet[5020$\pm$800~cm$^{-3}$]{2005ApJS..157..371R} and \citet[4500$^{+1100}_{-900}$~cm$^{-3}$]{2008ApJ...677.1100W}.

Table~\ref{tab:n6543_2} shows the derived values of the ionic and total abundances. He abundances are spread within the range from 0.098$\pm$0.002 to 0.124$\pm$0.024 for the different nebular components. \citet{2000MNRAS.318...77H} studied two regions of this nebula (named \textit{east} and \textit{north}) and they found a similar helium abundance behaviour of 0.1 and 0.13, respectively, for these two regions. Though no errors are quoted for the latter results, authors argue that this dissimilitude gives room for a possible spatial variation in the nebula. 
For the O abundance, there is no trend among nebular components, and our results are in agreement with published values (see Fig.~\ref{fig:abundances2}), except for both LISs located along the position angle of 5$^{\circ}$. The latter have slightly lower oxygen abundances, 
which are actually on the limit to, within the errors, agree with that of the other components. As far as N/H is concerned, there could be a very small variation across the nebula, with the highest values found in the A and D rims, together with the J' LISs (all along the same PA, 5$^{\circ}$), possible variation that was also reported by \citet{1994ApJ...424..800B}. The Ne abundance also follows the possible small variation through the components (peaking at the position of the rims AA' and DD') and in average it is in good agreement with the literature. Considering the average of N/O and He/H, we conclude that this nebula is of non-Type I class. The total Ar abundances present some subtle variations, being the lowest in the DD' (rim) and invaluable for the LISs. For Neb, Ar/H is sub-estimated by one order of magnitude, when compared with some authors, but close to the values reported by \citet{2004MNRAS.349..793P}, as it can be seen in  Fig.~\ref{fig:abundances2}.

\subsection{NGC 6905}

\begin{figure}
    \centering
    \fbox{\includegraphics[width=0.7\columnwidth]{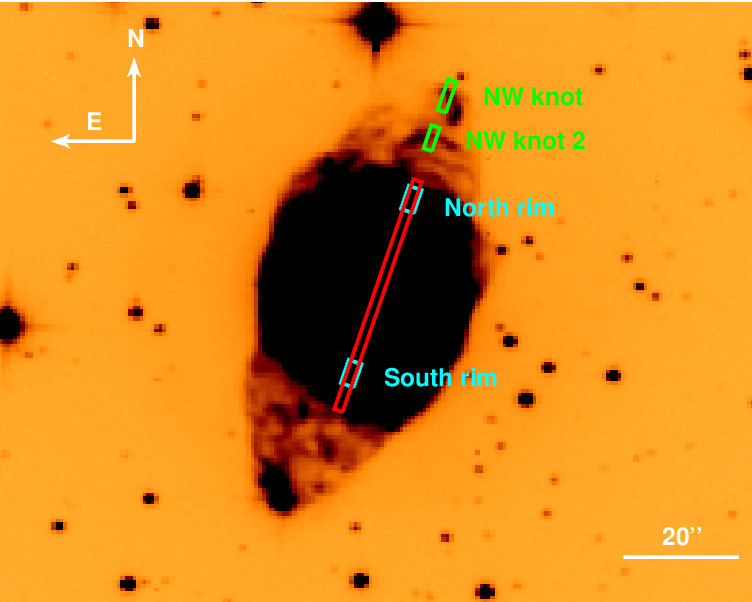}}
    \includegraphics[width=0.75\columnwidth]{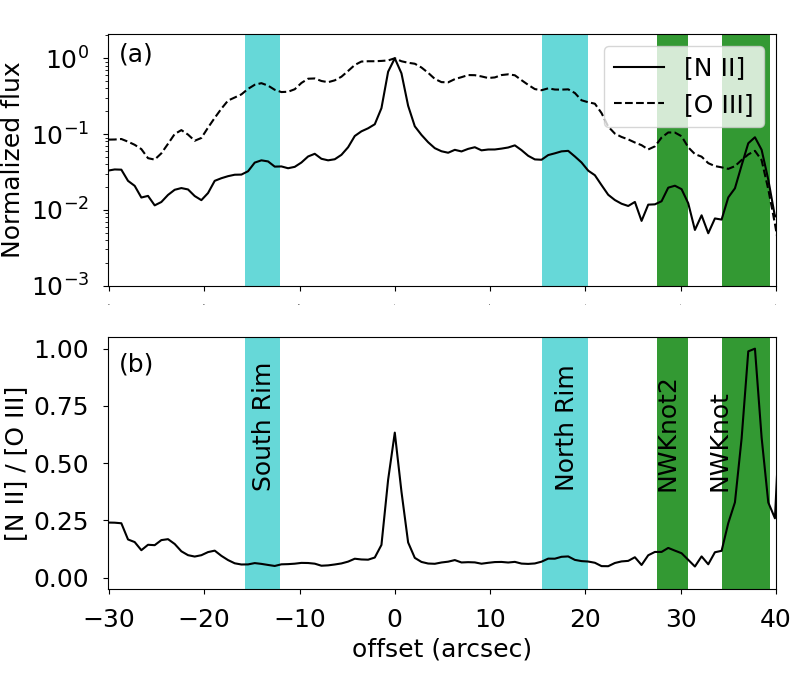}
    \caption{\textit{Upper panel:} H$\alpha+$[N~{\sc ii}] image of NGC 6905 from \citet{2003MNRAS.340..417C}. The size of the field is 130$\times$104 arcsec$^{2}$. The nebular components under analysis are indicated by the boxes: the nebular region (Neb, 42 arcsec, in red); the rims (4.2 arcsec for south and 5.6 arcsec for north one, in cyan); the LISs (5.6 arcsec for NW knot and 4.2 arcsec for NW knot2, in green). \textit{Lower panels:} Same as previous Figures.}
    \label{fig:n6905}
\end{figure}

\begin{figure*}
    \centering
    \includegraphics[width=0.64\textwidth]{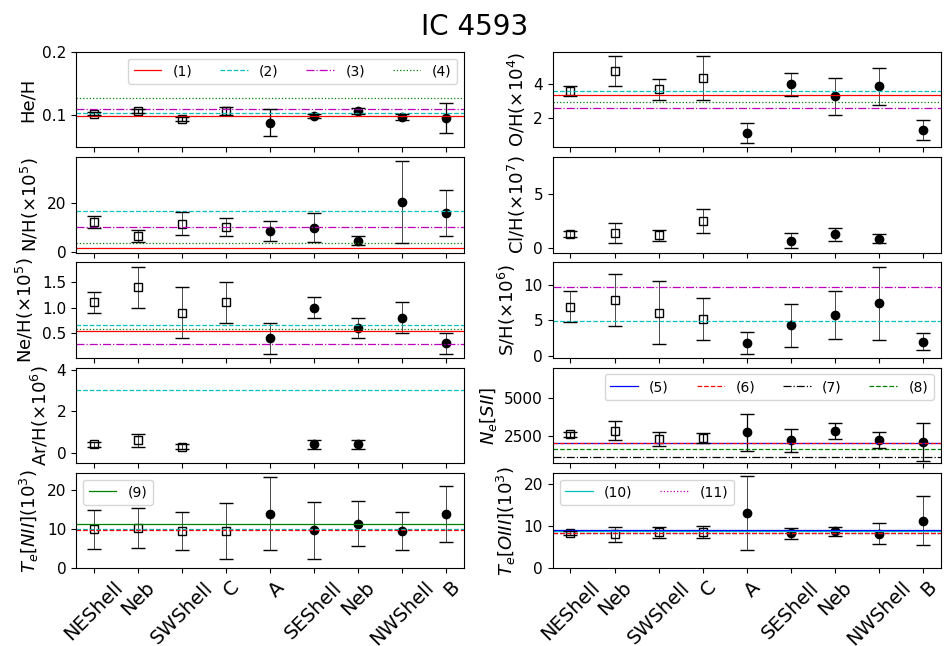}
    \includegraphics[width=0.64\textwidth]{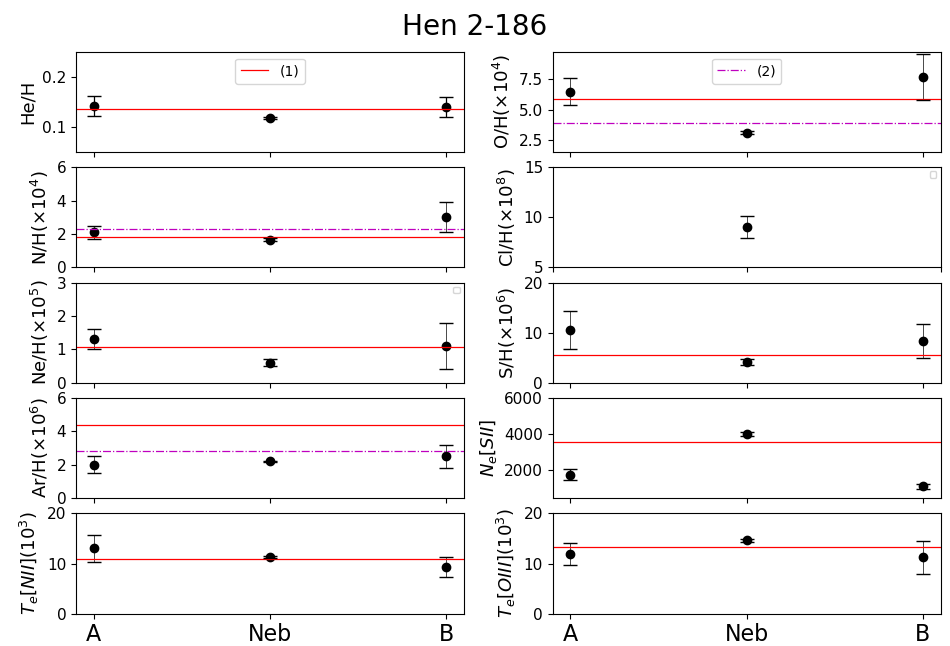}
    \includegraphics[width=0.64\textwidth]{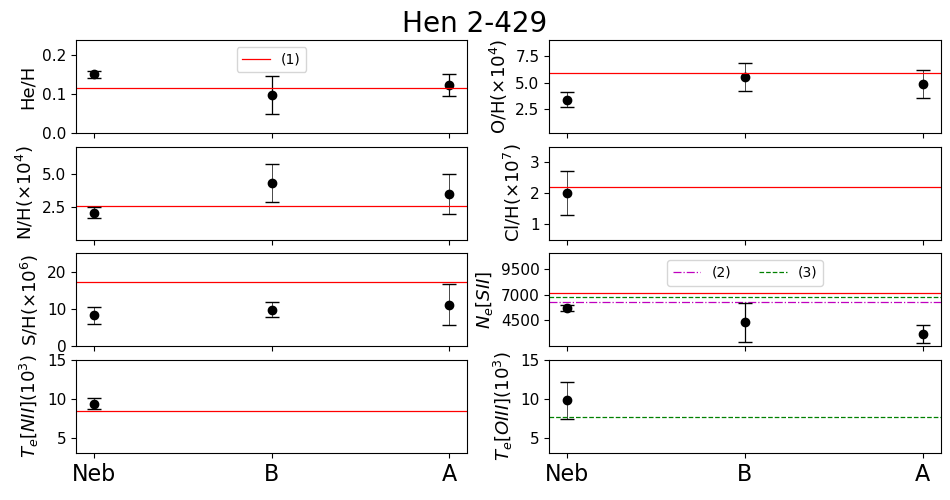}
    \caption{Variation of total abundances, electronic density and temperature across nebular components for IC 4593, Hen 2-186 and Hen 2-429. The horizontal lines represent values published in the literature, only for comparison with our results. When a studied nebula has more than one PA, different symbols represent different PAs.  References for IC 4593: (1) \citet[][]{1991ApJS...76..687P}, (2) \citet[][]{1996RMxAA..32...47B}, (3) \citet[][]{1998ApJ...493..247K}, (4) \citet[][]{2006ApJ...651..898S}, (5) \citet[][]{1978ApJ...219..914B}, (6) \citet[][]{2009ApJ...694.1335D}, (7) \citet[][]{1989ApJ...343..811S}, (8) \citet[][]{1998A&A...340..527P}, (9) \citet[][]{1986ApJ...308..322K}, (10) \citet[][]{1978ApJ...225..527K}, (11) \citet[][]{2005ApJS..157..371R}. References for Hen 2-186: (1) \citet[][]{2010RMxAA..46..159C}, (2) \citet[][]{2017MNRAS.471.4648V}. References for Hen 2-429: (1) \citet[][]{2007A&A...463..265G}, (2) \citet[][]{2000A&A...362.1008G}, (3) \citet[][]{2006RMxAA..42...53M}. }
    \label{fig:abundances1}
\end{figure*}

This nebula has a bright, broadly spheroidal shell, with a roughly conical shape extending over $\sim$82~arcsec, with a pair of LISs located $\sim$35'' from the central star. The H$\alpha$+[N~{\sc ii}] images from \citet{2003MNRAS.340..417C} reveal this pair of LISs, and also another one near the NW LIS, which we refer to as NW~knot~2. The opposite knot of the pair (SE knot) is close enough to a field star to be contaminated, so it was excluded from our spectroscopic analysis. The nebular components that we selected for analysis are shown in Fig.~\ref{fig:n6905}, while the results of their analysis appear in 
Table~\ref{tab:n6905_1}.

The interstellar extinction coefficient of NGC~6905 does not shown significant variation 
from one to another nebular component, with an average of 0.05$\pm$0.02.
Our c$_{\beta}$ agrees with 
one of the previous works, 0.11$\pm$0.18 \citep[][]{1976AJ.....81..407C}. Other authors reported extinctions of 0.27 \citep[][for a PA of 90$^{\circ}$]{1994MNRAS.271..257K} or 0.23, 0.22, 0.29 \citep[][three different slit positions]{1998A&A...337..866P}, 0.55 \citep[][]{1970ApJ...160..887K} and 0.93 \citep[][]{1986ApJ...308..322K}. In fact, recently \citet[][]{2022MNRAS.509..974G} studied almost the same low- and high-ionization structures as in the present work (and named A1, A2 and A3 our NWknot, NWknot2 and north rim, as well as A6 for our south rim), by using longslit spectra with a PA of 155$^{\circ}$. These authors also obtain a similar c$_{\beta}$ variation, from 0.05 to 0.23.

\begin{figure*}
    \centering
    \includegraphics[width=0.64\textwidth]{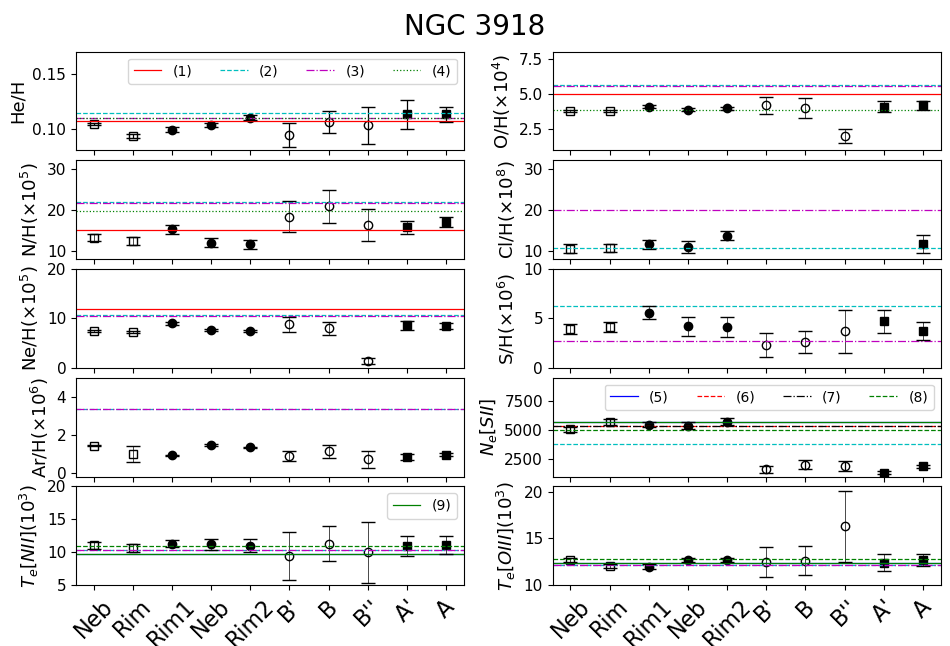}
    \includegraphics[width=0.64\textwidth]{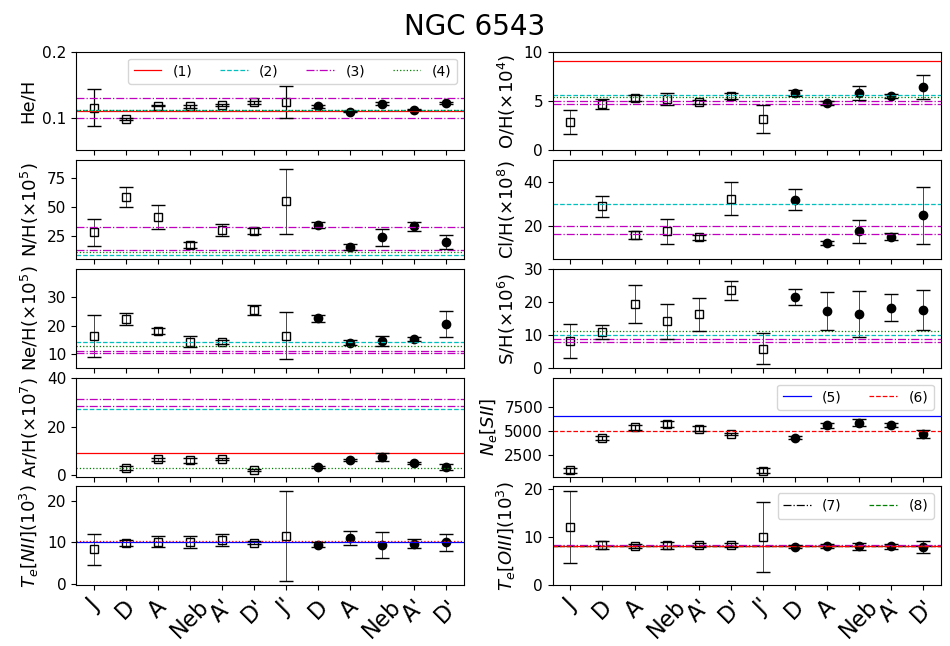}
    \includegraphics[width=0.64\textwidth]{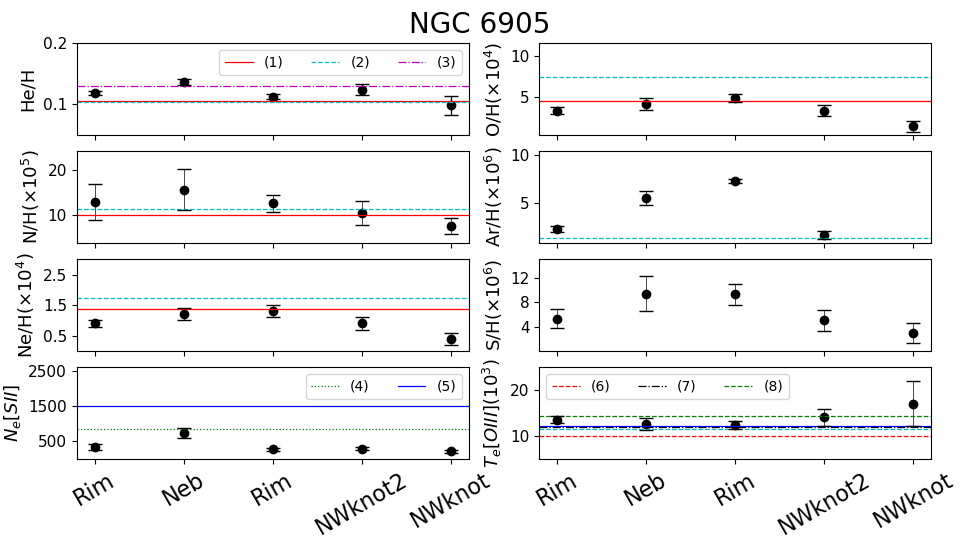}
    \caption{Same as Fig.~\ref{fig:abundances1} for NGC 3918, NGC 6543 and NGC 6905. References for NGC 3918: (1) \citet[][]{1991ApJS...76..687P}, (2) \citet[][]{2003PASP..115...80K}, (3) \citet[][]{2004AJ....127.2284H}, (4) \citet[][]{2018MNRAS.473..241H}, (5) \citet[][]{1987ApJ...314..551C}, (6) \citet[][]{1989ApJ...343..811S}, (7) \citet[][]{1998A&A...340..527P}, (8) \citet[][]{2015MNRAS.452.2606G}, (9) \citet[][]{1991ApJ...377..210R}. References for NGC 6543: (1) \citet[][]{1970ApJ...160..887K}, (2) \citet[][]{1999A&A...347..967P}, (3) \citet[][]{2000MNRAS.318...77H}, (4) \citet[][]{2004MNRAS.349..793P}, (5) \citet[][]{2004MNRAS.351.1026W}, (6) \citet[][]{2008ApJ...677.1100W}, (7) \citet[][]{1998A&A...340..527P}, (8) \citet[][]{2005ApJS..157..371R}. References for NGC 6905: (1) \citet[][]{1991ApJS...76..687P}, (2) \citet[][]{1994MNRAS.271..257K}, (3) \citet[][]{1995A&A...302..211S}, (4) \citet[][]{1989ApJ...343..811S}, (5) \citet[][]{1998A&A...337..866P}, (6) \citet[][]{1976AJ.....81..407C}, (7) \citet[][]{1986ApJ...308..322K}, (8) \citet[][]{1970ApJ...160..887K}. }
    \label{fig:abundances2}
\end{figure*}

We did not find significant difference in T$_{e}$[O~{\sc iii}] among the structures, being the mean 13960$\pm$1088~K. This temperature is higher than some previous measurements (see Fig.~\ref{fig:abundances2}), but close to that of \citet[14300~K]{1970ApJ...160..887K}. 
\citet[][]{2022MNRAS.509..974G} obtained T$_{e}$[O~{\sc iii}] of $12840\pm1010$~K for A1, $12430\pm800$~K for A2, $12030\pm380$~K for A3 and $12930\pm440$~K for A6, all comparable with ours 
(nevertheless the central value we found for NWknot is a factor $\sim$1.3 higher). 
\citet{1998A&A...337..866P} also studied this nebula using three different positions angles: (i) 90$^{\circ}$, passing through the central star; (ii) 0$^{\circ}$, with 5'' offset to the north of the central star; and (iii) 90$^{\circ}$ for the SE LIS. Their T$_{e}$[O~{\sc iii}] of (12100$\pm$800)~K, (13100$\pm$800)~K and (14000$\pm$3000)~K for each of these slits, respectively, are not in disagreement with our mean T$_{e}$[O~{\sc iii}]. Notice that a spatial variation in T$_{e}$[O~{\sc iii}] may be indicated from the T$_{e}$[O~{\sc iii}] of \citet{1998A&A...337..866P}, but the uncertainties do not allow for a robust conclusion. An analysis of c$_{\beta}$ and T$_{e}$ with IFU will be useful to further investigate the possibility of their spatial variation in NGC~6905. As for the electron density, the Neb value we derived from the [S~{\sc ii}] doublet is (740$\pm$150)~cm$^{-3}$, whereas densities of the rims are much lower having values of 350$\pm$80 and 280$\pm$40, and LISs are less dense by a factor of $\sim$2.5 and $\sim$3.4. Yet, the Neb density agrees, within the errors, with \citet[850~cm$^{-3}$]{1989ApJ...343..811S}, but it is lower than that of \citet[1500~cm$^{-3}$]{1998A&A...337..866P}. 
Regarding the LIS, despite the fact that both sulfur diagnostic lines were detected, the line ratio lies outside the theoretical curve \citep[e.g.][]{2006agna.book.....O,1989ApJ...343..811S}, indicating a very low electron density. Thus, we argue for N$_{e}$[S~{\sc ii}]$<$220~cm$^{-3}$ for NW~knot. Accounting for the errors, \citet[][]{2022MNRAS.509..974G} LISs ($500\pm290$~cm$^{-3}$) and north rim ($480\pm220$~cm$^{-3}$) densities are also in agreement with our results.

NGC~6905 abundances are listed in Table~\ref{tab:n6905_2}. As expected, there is no variation in He abundance from one to another component, being He/H=$0.118\pm0.038$, the average value, in agreement with previous literature reports (see Fig.~\ref{fig:abundances2}). It should be noted that a strong variation in the He~{\sc ii}~$\lambda$4686/H$\beta$ has been reported by \citet{1961SvA.....5..186V}, with an intensity ratio varying from 0.5 to 1.3 over $\sim$15 years (between 1945 and 1959). In our analysis we find, for Neb, a He~{\sc ii}~$\lambda$4686/H$\beta$ ratio of $\sim$1.05.  \citet[][]{2022MNRAS.509..974G} succeed in separating the central star  
from the nebular emission. 
Ours and theirs He~{\sc ii}~4686~\AA~ for both structures match well, apparently indicating both emissions are blended. 
With the exception of the NW knot, for which we find an under-abundance of a factor $\sim$2.8 respect to the Neb, the O/H does not change among the components. 
Per structure, N/H are higher than those reported by \citet[][]{2022MNRAS.509..974G}, which may be due to need of applying the T$_{e}$[O~{\sc iii}] instead of T$_{e}$[N~{\sc ii}] to derive the nitrogen abundance. On one hand, N/O is similar throughout the nebular components, but, on the other hand, higher in the NW knot by a factor $\sim$1.4. Considering He/H and N/O in Table~\ref{tab:n6905_2}, we can conclude that NGC~6905 classifies as Type I.
The total Ne/H abundances are also similar among the nebular components and in agreement with the values reported in literature. On the other hand, Ar/H presents a variation with the higher value found for one of the rims (also higher than in the literature, as in Fig.~\ref{fig:abundances2}) but not measurable for NW knot. S/H is found to be nearly constant among the nebular components except the NW LIS, for which it is lower by a factor of $\sim$3.2 respect to the Neb, difference that was also reported by \citet[][]{2022MNRAS.509..974G}. As in the case of a couple of LISs previously mentioned, some X/H are lower in the LIS than in the rest of the components, as a consequence of the T$_{e}$[O{\sc iii}] that is also much higher (and highly uncertain).

\section{Discussion}

In the following we discuss different aspects of the physical, chemical and excitation properties of the six PNe whose results were given above, focusing on their variation across slits and/or components, and also allowing for the comparison with the literature. 

\subsection{Extinction variation}
For two of the PNe in the sample we found variations in the extinction coefficient. The first case is IC~4593, for which this coefficient is ranging from $0.02\pm0.02$ to $0.22\pm0.01$. We also want to note that the intensities of H$\gamma$ and H$\delta$ lines display a small variation from one another structure, suggesting a real internal spatial variation in this coefficient. The second case is NGC~6543, which has a c$_{\beta}$ that takes values from $0.02\pm0.02$ to $0.26\pm0.01$. For this nebula, there is a PA in which the H$\alpha$ line was saturated, therefore c$_{\beta}$ was estimated using only H$\gamma$ and H$\delta$. Even so, the other PA studied, whose c$_{\beta}$ were calculated also using H$\alpha$, presents as well extinction coefficient variations. As with the previous nebula, we note that the intensity of H$\gamma$ and H$\delta$ lines show small variation. In addition, in recent works of \citet[][]{Akras2020} and \citet[][]{2022MNRAS.512.2202A}, in which longslit were simulated on IFU data, the authors verify that variations in the extinction coefficient are present even as a result of only changing the PA. Therefore, these spatial variations in the extinction coefficient, in both nebulae, are likely real, and they can only be much better studied through IFU data.

\subsection{Densities and temperatures}

\begin{figure}
    \centering
    \includegraphics[width=\columnwidth]{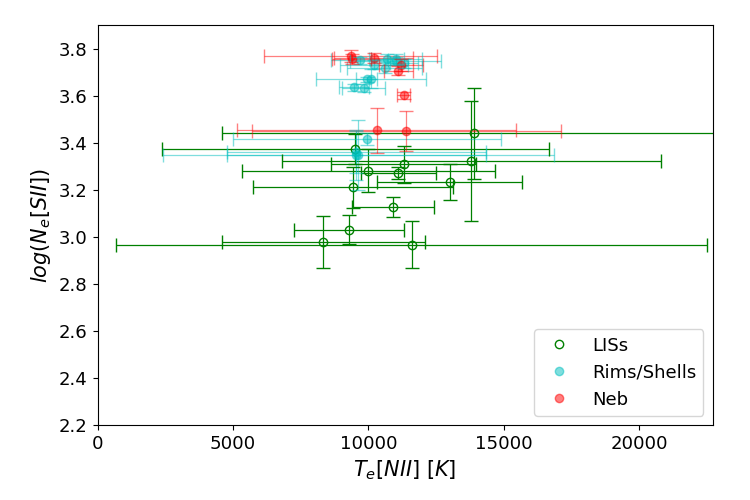}
    \includegraphics[width=\columnwidth]{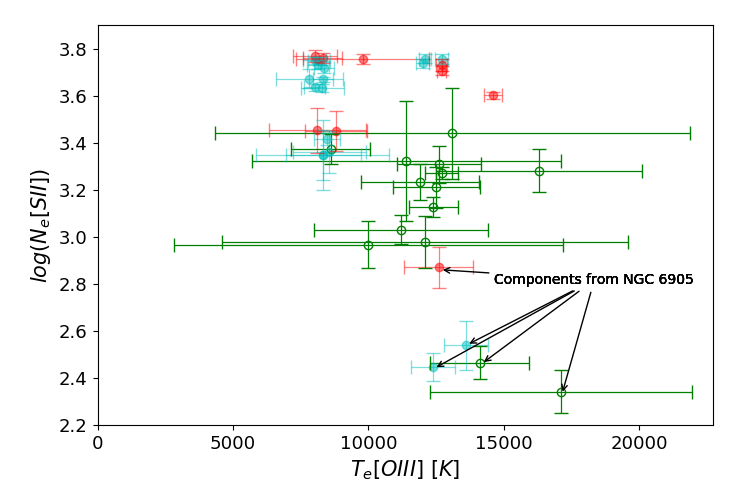}
    \caption{T$_{e}$ versus N$_{e}$ for all the components in our PNe's sample. \textit{Upper panel}: T$_{e}$ using [N~{\sc ii}] diagnostic lines versus log(N$_{e}$[S~{\sc ii}]). \textit{Lower panel}: T$_{e}$ using [O~{\sc iii}] diagnostic lines versus log(N$_{e}$[S~{\sc ii}]). Arrows are indicating these quantities for NGC 6905, which it was not possible to determine T$_{e}$[N~{\sc ii}].}
    \label{fig:T_N}
\end{figure}

Figure~\ref{fig:T_N} displays 
N$_{e}$[S~{\sc ii}] 
as a function of T$_{e}$ from the [N~{\sc ii}] (upper panel) and the [O~{\sc iii}] (lower panel) lines. 
The main conclusions drawn from these plots are: (i) T$_{e}$[N~{\sc ii}] is almost invariant through the components considering their uncertainties, whereas LISs' T$_{e}$[O~{\sc iii}] is higher compared to the rims/shells, although the large uncertainties 
do not allow for a robust conclusion; and (ii) N$_{e}$[S~{\sc ii}] is generally lower for LISs relative to the rest of the components.

\begin{figure}
    \centering
    \includegraphics[width=\columnwidth]{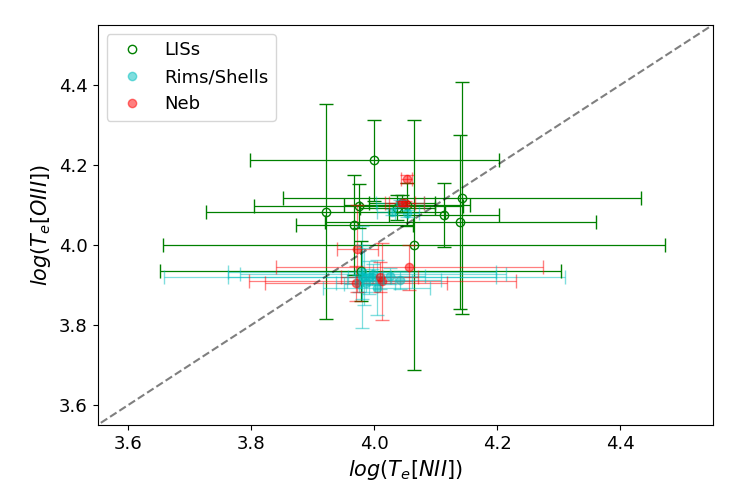}
    \caption{T$_{e}$[N~{\sc ii}] versus T$_{e}$[O~{\sc iii}] for all the components in our PNe's sample. The dashed gray line represents the identity.}
    \label{fig:T_T}
\end{figure}

The comparison between the two T$_{e}$ indicators -- from [N~{\sc ii}] and [O~{\sc iii}] lines -- 
is presented in Fig.~\ref{fig:T_T}. In particular, we find that the mean value for T$_{e}$[N~{\sc ii}] 
is 11000$\pm$1600 and T$_{e}$[O~{\sc iii}] is 12600$\pm$1200~K for the group of LISs, whereas for the group of rims/shells these values are 10200$\pm$800 and 9510$\pm$220~K, respectively. 
This difference in T$_{e}$ between the two indicators is also present for the sample of PNe  discussed by \citet[][]{2006agna.book.....O} (see their Fig.5.3) and attributed to excitation differences among the nebulae.

\subsection{Abundance trends}

\begin{figure}
    \centering
    \includegraphics[width=\columnwidth]{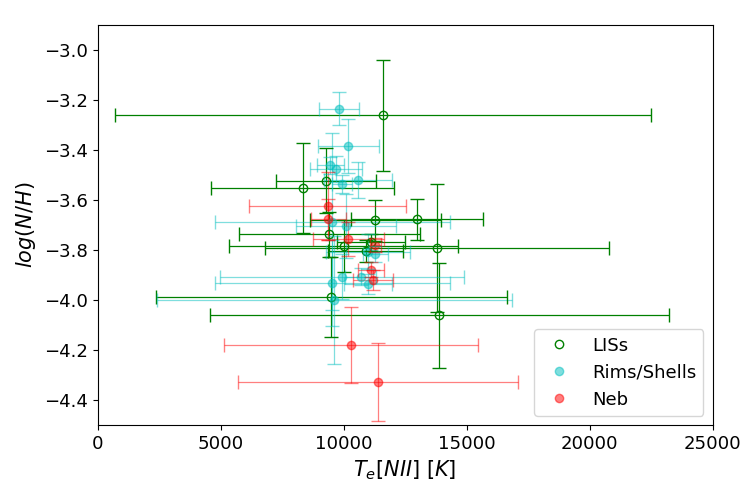}
    \includegraphics[width=\columnwidth]{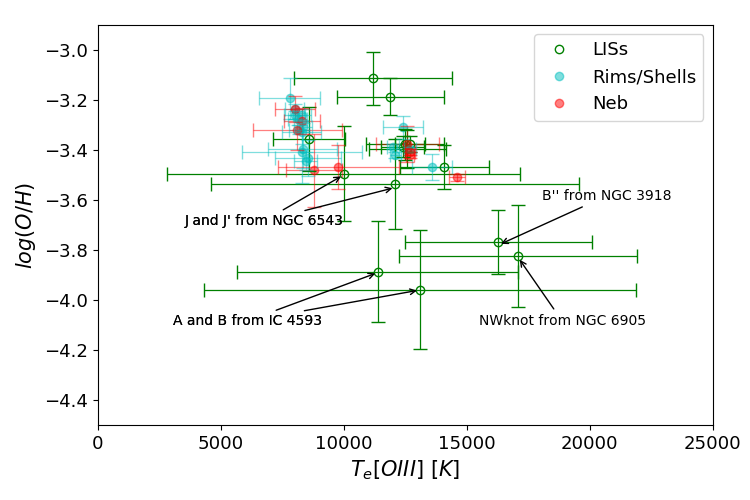}
    \caption{Variation of oxygen and nitrogen abundances as a function of electronic temperature. \textit{Upper panel:} T$_{e}$[N~{\sc ii}] versus log(N/H), in which there is no tendency considering the errors. \textit{Lower panel:} T$_{e}$[O~{\sc iii}] versus log(O/H). Arrows are indicating the position of the LISs having lower O/H than the rest of the points (see text).}
    \label{fig:T_X}
\end{figure}

Fig.~\ref{fig:T_X} shows that the range of log(N/H) and log(O/H) we derived are $-$4.4 to $-$3.0 and $-$4.2 to $-$3.0, respectively. A negligible variation in log(N/H), for T$_{e}$[N~{\sc ii}], was observed, regardless the nebular components. On the contrary, the derived O abundances show a non-negligible variation with T$_{e}$[O~{\sc iii}]. In particular, most of log(O/H) lie within the range of $-$3.6 to $-$3.1. %
regardless of T$_{e}$[O~{\sc iii}]. 
But there are four cases with log(O/H)$<-$3.7: two LISs in IC~4593, one LIS in NGC~3918 and one in NGC~6905. The latter cases are also characterized by T$_{e}$[O~{\sc iii}]$>$15000~K, which is thus responsible for the resulting lower O abundances relative to the rest of the components in these PNe. Notice that all these cases are also described by very large uncertainties in T$_{e}$[O~{\sc iii}] due to the detection of weak [O~{\sc iii}]~4363\AA~emission line.

\begin{figure*}
    \centering
    \includegraphics[width=\columnwidth]{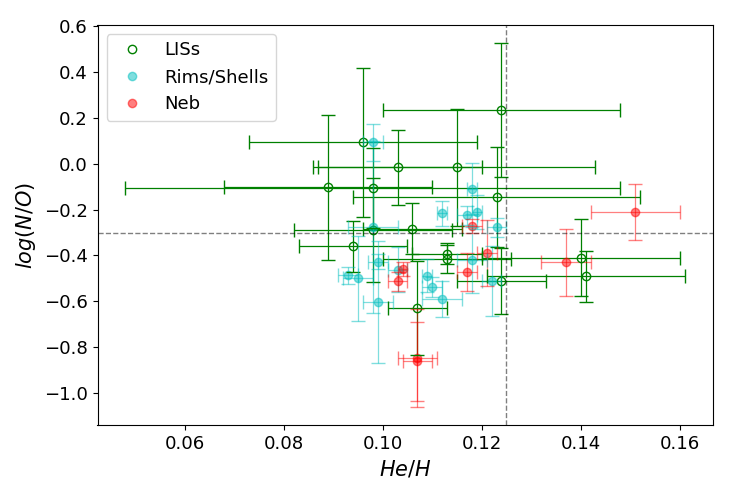}
    \includegraphics[width=\columnwidth]{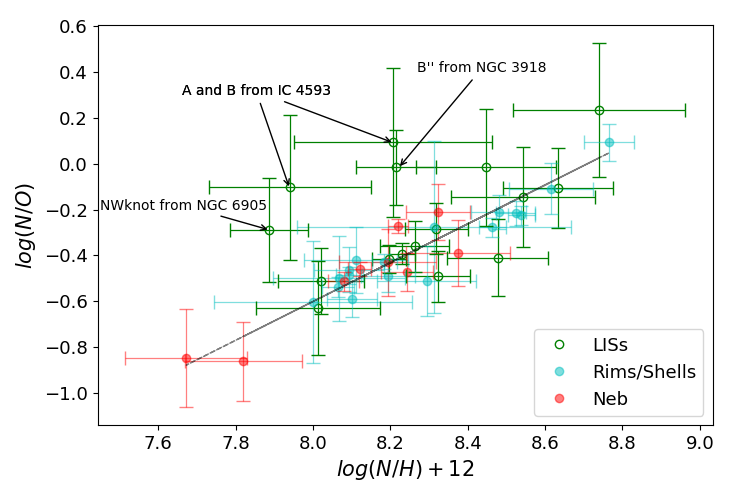}
    \includegraphics[width=\columnwidth]{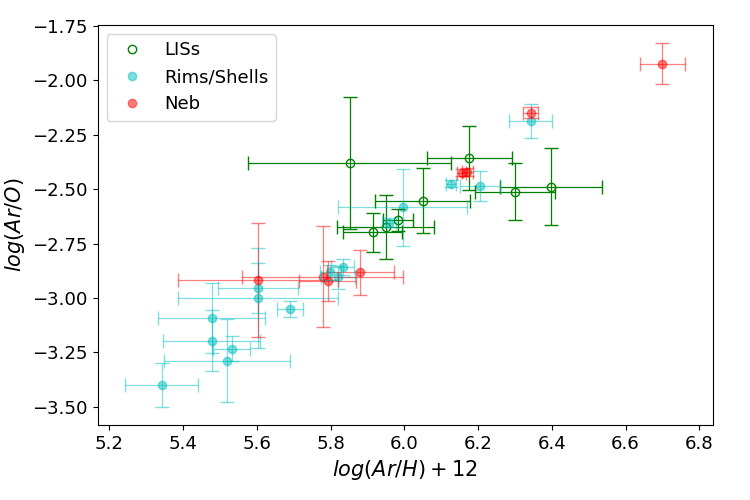}
    \includegraphics[width=\columnwidth]{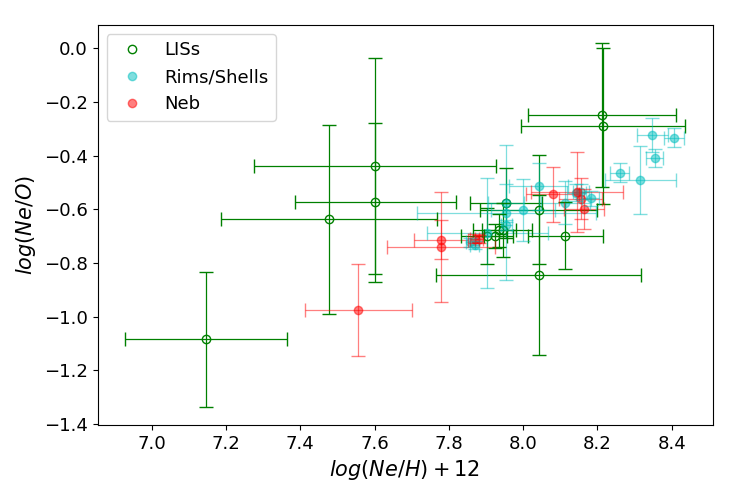}
    \includegraphics[width=\columnwidth]{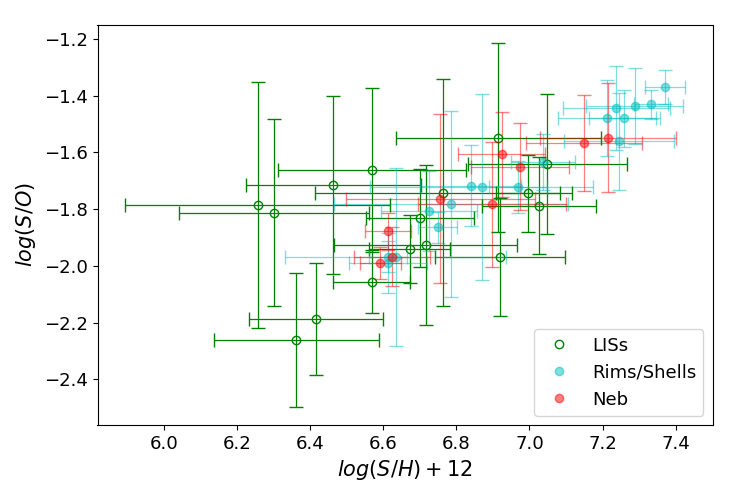}
    \includegraphics[width=\columnwidth]{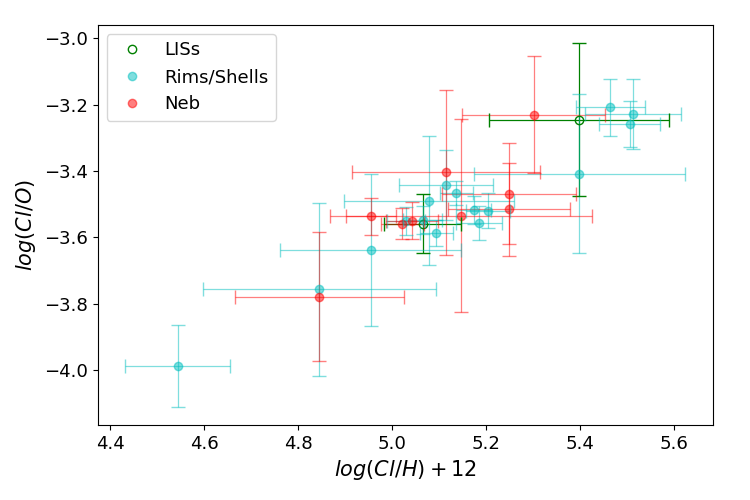}
    \caption{\textit{Upper panels}: N/O abundance ratio plotted against He abundance (left) for our PN sample. The dashed lines are delimiting the criteria giving by \citet[][]{1978IAUS...76..215P} to define Type I PNe. In the right panel, log(N/O) versus log(N/H)+12, where the dashed line is representing the best linear fit to our data (see text). \textit{Middle panels}: relationship between Ar/O (Ne/O) abundance ratio and the total Ar (Ne) abundance on the left (right). \textit{Lower panels}: relationship between S/O (Cl/O) abundance ratio and the total S (Cl) abundance on the left (right).}
    \label{fig:XH_XO}
\end{figure*}

Figure~\ref{fig:XH_XO} presents the correlation among various abundance 
ratios. The dashed lines in the He/H versus log(N/O) plot, define the limits of Type I PNe \citep[He/H$\geq$0.125 and log(N/O)$\geq-0.3$,][]{1978IAUS...76..215P}. All but one PNe (NGC 6905) are classified as non-Type I. The log(N/H)+12 versus log(N/O) plot gives the best linear relation as $log(N/O) = 0.85 (log(N/H)+12) - 7.37$, with a goodness-of-fit ($R^{2}$) equal to $0.83$, if eliminating the LISs marked with arrows (the same highlighted in Fig.~\ref{fig:T_X}). Our analysis returns a slightly different slope and intercept when compared to a sample of PNe with [WR]-type 
central stars \citep[0.73 and $-$6.50, ][]{2013A&A...558A.122G} or a sample of 
PNe with LISs (0.74 and $-$6.50, Paper~I).

The correlations between log(X/H)+12 and log(X/O) for X = Ar, Ne, S and Cl are also presented in 
Fig.~\ref{fig:XH_XO}. No significant contrast between LISs and the rims/shells are found for these chemical abundances, in agreement with the conclusions from Paper~I, and other previously published analyses.

\subsection{Central star properties}

\begin{table}
\caption{General characteristics of central star planetary nebulae (CSPNe) from \citet{2020A&A...640A..10W}.}
\label{tab:cspn}
\begin{center}
    \begin{tabular}{lcccc}
\hline
          & Type       & T$\rm _{eff}$ {[}K{]} & L [L$_{\odot}$] & log(g) \\
\hline          
IC 4593   & O(H)5f             & $\sim$48000       & $\sim$5500      & 3.70   \\
Hen 2-186 & Cont. $^{\dag}$    & $\sim$107500      & $\sim$7900      & 5.40   \\
Hen 2-429 & [WC 4] $^{(1)}$    & $\sim$23000 $^{\dag\dag}$   & $\sim$5900      & -      \\
NGC 3918  & O(H)               & $\sim$150000      & $\sim$5000      & 5.56   \\
NGC 6543  & Of-WR(H) $^{(2)}$  & $\sim$60400       & $\sim$3800      & 4.70   \\
NGC 6905  & [WO 2] $^{(3)}$    & $\sim$130900      & $\sim$10200     & 4.70   \\
\hline
\end{tabular}
\end{center}
Notes: $^{\dag}$ Type \textit{cont.} means that its spectrum has a high S/N ratio with no stellar features. $^{\dag\dag}$ Such a low temperature contradicts our spectroscopic data in which we detect the He~{\sc ii} $\lambda$4686 line. \\
$^{(1)}$ We detected emission lines from this type of CSPN (see Table~\ref{tab:h429_1}) having FWHM in good agreement with the literature \citet[][37$\pm$1]{2002AJ....124..464D}. \\
$^{(2)}$ We identify emission lines with a stellar origin (see Table~\ref{tab:n6543_1}). \\
$^{(3)}$ According to \citet{1968IAUS...34..339A} and \citet{Cuesta1993}, this nebula owns one of the most broadened O~{\sc vi} emission lines observed among PNe, which are also detected in our spectra (see Table~\ref{tab:n6905_1}) and reported by \citet[][]{2022MNRAS.509..974G}

\end{table}

Concerning the central stars of the PNe studied here (Table~\ref{tab:cspn}), we find that these PNe containing LISs have either [WR]- or O(H)-type \citep[see also,][]{Miszalski2009} covering a large range of T$_{\rm eff}$, $\sim$45000 K to $\sim$150000 K, L ranging from $\sim$3800 L$_{\odot}$ up to $\sim$10000 L$_{\odot}$, and log(g) taking values from 3.7 to 5.5.

\subsection{Photo- versus shock-excitation of LISs}

The long-standing problem of the excitation mechanism of LISs 
can also be addressed for our sample of PNe. In Paper~I, the $f_{\rm shocks}/f_{\rm star}$ ratio was defined to explore the contribution of shocks on LISs and their host PNe. 
This ratio describes the ionizing photon flux emitted from the central star ($f_{\rm star}$) and the ionizing photon flux produced by a potential shock interaction ($f_{\rm shocks}$). Paper~I authors 
concluded that log$(f_{\rm shocks}/f_{\rm star}) > -1$ defines the zone of shock-dominated structures, while log$(f_{\rm shocks}/f_{\rm star}) < -2$ encompasses 
the photoionization-dominated structures, and the transition between the two, $-2 < log(f_{\rm shocks}/f_{\rm star}) < -1$, is where both mechanisms are present.
Given the (generally uncertain) parameters entering this ratio, a deeper understanding of it is in place, before further dedicated analyses. 

The ionizing photon flux, $f_{\rm shocks}$, is determined from the total radiative flux, $F_{\rm shocks} = 2.28\times 10^{-3} (V_{\rm shock}/100km s^{-1})^{3} (N_{e}/cm^{-3})$ ergs~cm$^{-2}$~s$^{-1}$, where $V_{shock}$ is the shock velocity and $N_{e}$ is the density of the pre-shocked gas \citep[][]{1996ApJS..102..161D}, divided by the average energy of a photon emitted from the central star $(L/S_{*})$. Thus, the equation becomes $f_{\rm shocks}/f_{\rm star} = 9.12 \times 10^{-3} (V_{shock}/100kms^{-1})^{3} (N_{e}/cm^{-3}) (\pi d^{2}/L)$, where $d$ is the LISs' distance from the central star (in cm) and $L$ is the luminosity of the central star (in erg~s$^{-1}$) \citep[][]{2019MNRAS.489.2923L}. The issue here is that the usually uncertain distances to PNe ($\propto$~d$^{2}$), the unknown density of pre-shocked gas ($\propto$~N$_{\rm pre-shock}$) and the exact velocity of the shock wave ($\propto$~V$_{\rm shock}^3$)  
prevent a unequivocal determination of the $f_{\rm shocks}/f_{\rm star}$ ratio. The explicit contribution of these parameters is as below. 

\noindent \emph{(i) Distances} - 
The distances of the PNe in our sample were mainly adopted from \citet{Frew2016,Guerrero2020}, otherwise additional references are quoted. 

\noindent \emph{(ii) Pre-shock density} - 
Because there is very little known about the pre-shock density, the $f_{\rm shocks}/f_{\rm star}$ 
ratio is determined for a range of N$_{\rm pre-shock}$ from 0.1 to 10000 cm$^{-3}$. We also point out that the density of the post-shocked gas (LISs) and pre-shocked gas (surrounding gas) are dependent parameters. At the shock jump, contact discontinuity, 
we get N$_{\rm post-shock}$ = (($\gamma$+1)$M^2$)/(($\gamma$-1)$M^2$+2)$\times$N$_{\rm pre-shock}$ \citep[see e.g.][]{2004A&A...416..623M}, where $\gamma$ is the adiabatic index (5/3 for an ideal gas) and M is the Mach number of the shock wave, defined as V$_{\rm shock}$/V$_{\rm sound}$, where V$_{\rm sound}$ is the sound speed in the pre-shocked gas and its value is $\sim$12-15~km~s$^{-1}$ for a gas with T$\sim$10000~K. For the case of a strong shock (M$\gg$1) and  ideal monoatomic gas, N$_{\rm post-shock} \approx 4\times$N$_{\rm pre-shock}$. The post-shocked gas becomes less dense  for lower Mach numbers (or V$_{\rm shock}$). In particular V$_{\rm shock}$~$>$45~km~s$^{-1}$ yields N$_{\rm post-shock}$/N$_{\rm pre-shock}$~$>$3, and 25$<$V$_{\rm shock}$~$<$45~km~s$^{-1}$ yields N$_{\rm post-shock}$/N$_{\rm pre-shock}$~$>$2 (for V$_{\rm sound}$=15~km~s$^{-1}$). Therefore, 
$f_{\rm shocks}/f_{\rm star}$ ratio is computed considering a N$_{\rm pre-shock}$ that is consistent with the observed N$_{\rm post-shock}$ of LISs.

\begin{figure*}
    \centering
    \includegraphics[width=\columnwidth]{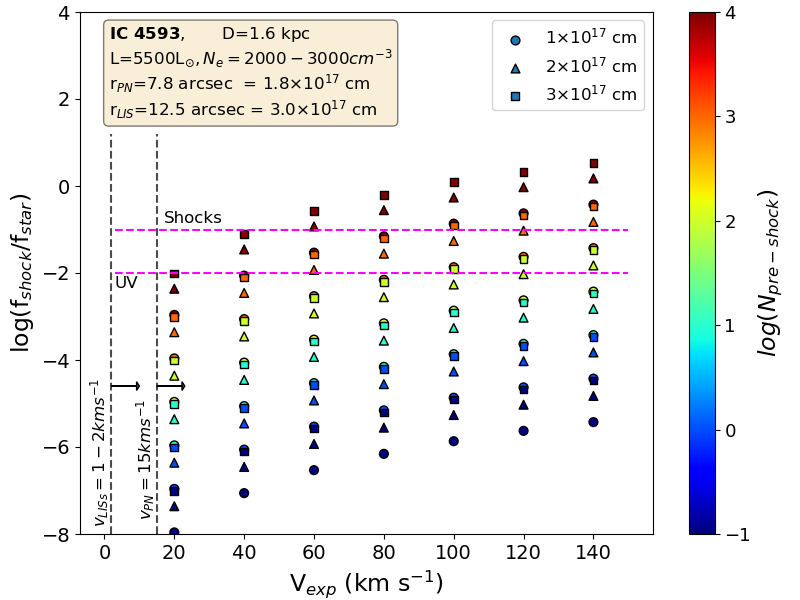}
    \includegraphics[width=\columnwidth]{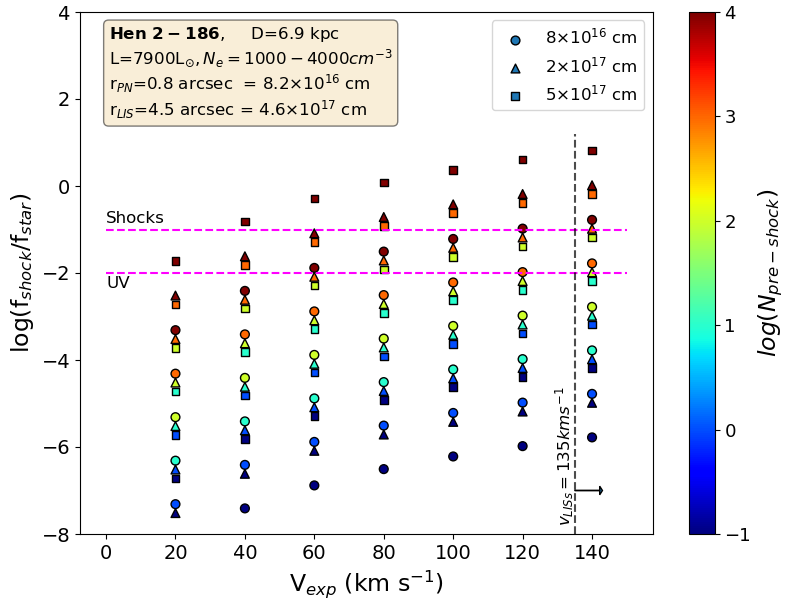}
    \includegraphics[width=\columnwidth]{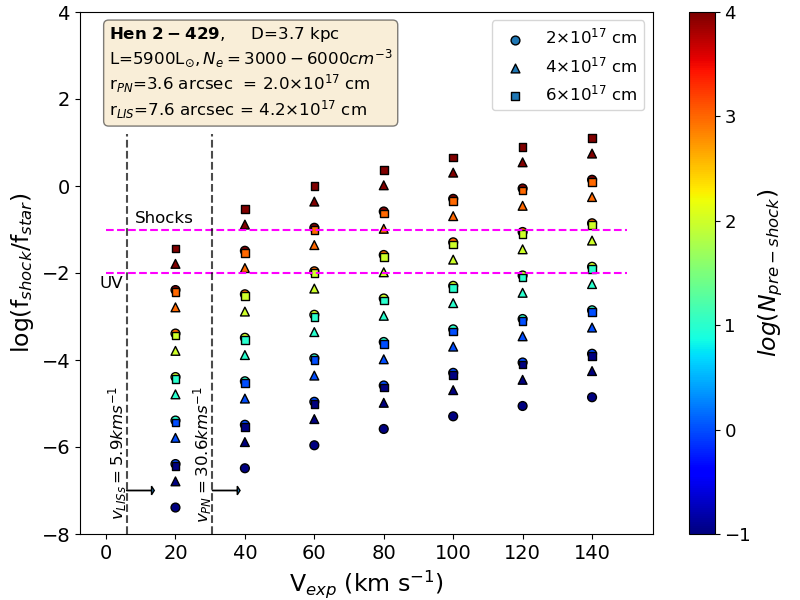}
    \includegraphics[width=\columnwidth]{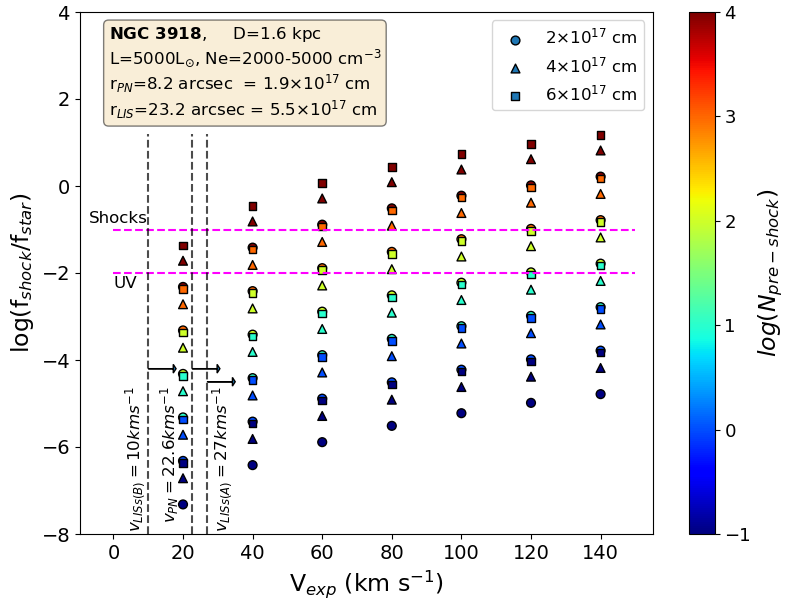}
    \includegraphics[width=\columnwidth]{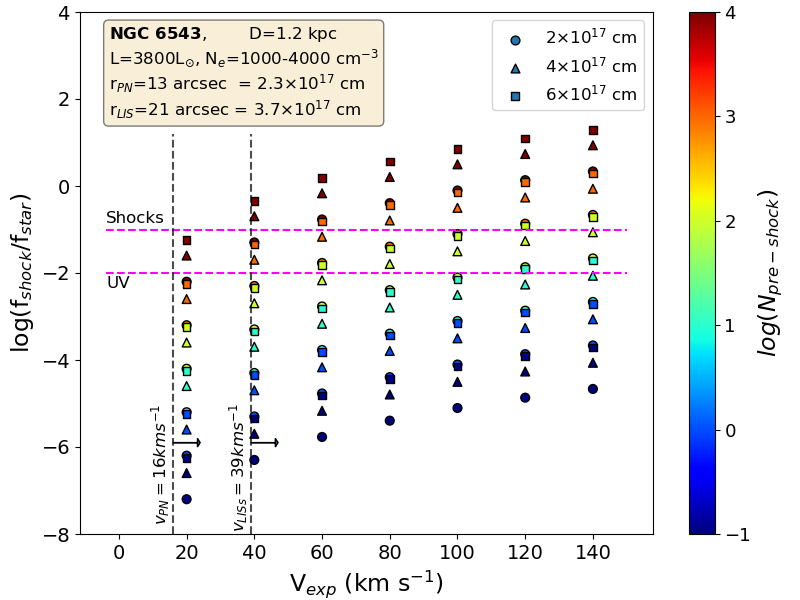}
    \includegraphics[width=\columnwidth]{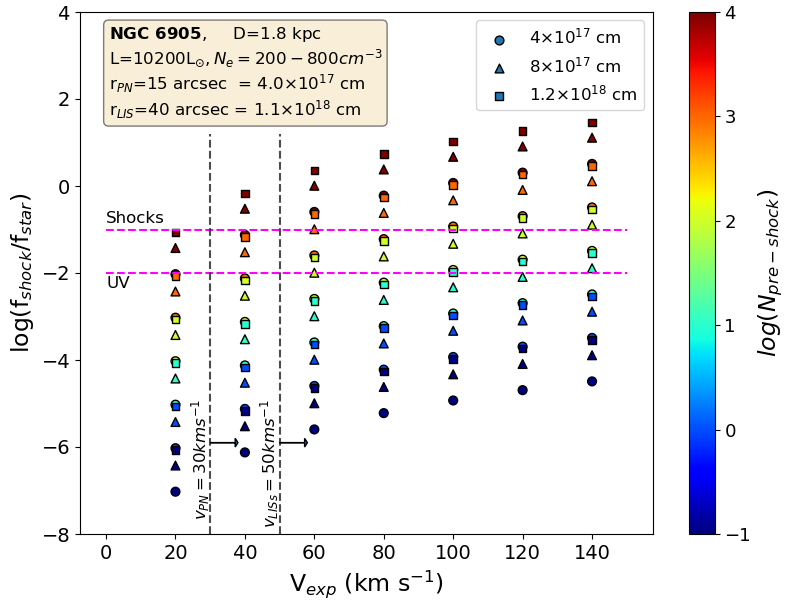}
    \caption{Variation of the $log(f_{\rm shocks}/f_{\rm star})$ ratio as a function of the shock velocity, for three distance from the central stars (circle, triangle and square symbols), and a range of pre-shock densities from 0.1 up to 10000~cm$^{-1}$ for each nebula of our sample. The velocities are taken from \citet[][]{Guerrero2020} for all PNe except NGC 6905 \citet{Cuesta1993}. The parameters of the central stars are taken from Table~\ref{tab:cspn}. The vertical dashed-line indicate the projected spectroscopic velocities of the LISs and host PNe. The horizontal arrows denote the lower limit of the observed velocities. The horizontal dashed-lines indicate the limits of the photo-ionization and shock dominated regions defined by \citet[][]{2016MNRAS.455..930A}.} 
    \label{fig:dds}
\end{figure*}

\noindent \emph{(iii) Inclination} - 
The limited information about the inclination of PNe and LISs does not allow for a direct determination of the expansion velocities. Hence, to derive the $f_{\rm shocks}/f_{\rm star}$ ratio 
a range of 
velocities, 20 to 140 km~s$^{-1}$, with steps of 20~km~s$^{-1}$, is considered. The 
radial velocities 
were adopted from \citet[][]{Guerrero2020} and \citet[][]{Cuesta1993}, and they are illustrated in Fig.~\ref{fig:dds} as lower limits (vertical dashed lines).

In an attempt of applying these ideas to the LISs of each PN in our sample, Fig.~\ref{fig:dds} displays the log($f_{\rm shocks}/f_{\rm star}$) ratio as function of the expansion velocity, for different distances from the central star (circle, triangle and square points) and various pre-shock densities (colour-bars).

\begin{itemize}
\item IC 4593 (top-left panel): 
The very small radial  
velocity of the LISs  (1$-$2~km~s$^{-1}$) 
suggests a highly inclined nebula,  which gives a 
expansion velocity up to 100 km~s$^{-1}$. These LISs have N$_{e}$ of $\sim$ 2000-3000 cm$^{-3}$ corresponding to N$_{\rm pre-shock}$ of $\sim$ 500$-$750~cm$^{-3}$. 
We argue that shock interaction becomes important for V$_{\rm exp}$~$>$60~km~s$^{-1}$. Lower pre-shock densities would require higher shock velocities. The minimum inclination angle with respect to the light of sight, for an expansion velocity of 60~km~s$^{-1}$ is $\sim$88$^{\circ}$. The contribution of shocks in the excitation of the LISs of IC~4593 is as likely as that of pure photo-ionization processes.

\item Hen 2-186 (top-right panel): This nebula shows the fastest LISs in our sample,  with V$_{\rm exp}$~$\gtrsim$ 135~km~s$^{-1}$.     
Such a high velocity is very likely responsible for shocked-heated gas. 
We get $-$1<log($f_{\rm shocks}/f_{\rm star}$)<$-$2 (in the transition zone), for 2$<$log(N$_{\rm pre-shock}$)$<$3, and 
higher ratios for higher densities. 
N$_{\rm pre-shock}$~$\sim$250$-$450~cm$^{-3}$ are implied from the LISs electron densities, $\sim$ 1000-1700 cm$^{-3}$. There is strong indication for a non-negligible contribution of shocks for the LISs of Hen 2-186.

\item Hen 2-429 (middle-left panel): LISs embedded in this nebula have N$_{e}$ in the range of $\sim$ 3000-4000~cm$^{-3}$, which corresponds to log(N$_{\rm pre-shock})$ between 2.8 and 3.0. The radial 
velocity of the LISs is 5.9~km~s$^{-1}$.
Only V$_{\rm exp} \gtrsim 40$~km~s$^{-1}$ (inclination angle of $\sim$81$^{\circ}$-82$^{\circ}$ respect to the light of sight) would be indicative of significant shocks contribution, even though, shock-excitation in the spectrum of the LISs of Hen~2-429 cannot be ruled out.

\item NGC 3918 (middle-right panel): An electron density of $\sim$1500$-$2000~cm$^{-3}$ has been estimated for its LISs that correspond to N$_{\rm pre-shock} \sim 400-500$ ~cm$^{-3}$,  depending on the Mach number. For such pre-shock densities only V$_{\rm exp}$~$>$40~km~s$^{-1}$ yield to a non-negligible contribution of shocks, at a distance from the central star of 6$\times$10$^{-17}$~cm (square symbols). The radial 
velocities of the LISs in NGC~3918 are 27~km~s$^{-1}$ (A) and 10~km~s$^{-1}$ (B). 
The excitation of both LISs can be attributed to shocks if V$_{\rm exp}$~$>$60~km~s$^{-1}$, or equivalently, their inclination angles respect to the light of sight are at least of 63$^{\circ}$ and 80$^{\circ}$, respectively.

\item NGC 6543 (bottom-left panel): In this nebula the LISs are characterized by $N_{e}\sim 950-1000$~cm$^{-3}$, or equivalently N$_{\rm pre-shock} \sim 250$~cm$^{-3}$. Their 
radial velocity is 39~km~s$^{-1}$ \citep{Miranda1992,Reed1999,Guerrero2020}. A distance of 1.2~kpc \citep[][]{GomezGordillo2020} 
to NGC~6543 is considered. 
The contribution of shocks becomes noteworthy for V$_{\rm exp}$~$>$60~km~s$^{-1}$. The morpho-kinematic analysis of NGC~6543 gives an inclination angle of $\sim$55 \citep{Miranda1992} implying a de-projected expansion velocity of $\sim$68~km~s$^{-1}$. The 
LISs excitation ratio in NGC~6543 is very likely between $-$2 and $-$1, in the transition zone, and therefore shock contribution cannot be ruled out.

\item NGC 6905 (bottom-right panel): 
It is characterized by very low LIS' N$_{e}$, between
200 and 300~cm$^{-3}$, or N$_{\rm pre-shock} \sim 55-70~cm^{-3} \approx 100~cm^{-3}$. The radial 
velocity of the PN and LIS are 30 and 50~km~s$^{-1}$, respectively. 
For such low N$_{\rm pre-shock}$, we obtain ratios greater than 
$-$2, for V$_{\rm exp}$ between 60 and 80~km~s$^{-1}$, or inclination angles of  45$^{\circ}$-65$^{\circ}$ with respect to the light of sight. Both mechanisms, photo-ionization and shock-heating likely contribute to the LISs' excitation in this PN.

\end{itemize}

\noindent Overall, taking into account the radial velocities, the LISs distances to the central stars, as well as the pre- and post-shock densities, we conclude that shock contribution cannot be ruled out as a possible excitation mechanism for most of the LISs in our sample. However, only the LISs in Hen~2-186, with radial 
velocity of 135~km~s$^{-1}$, unequivocally lie in the shock-dominated zone. We also note that pre-shock densities lower than 100~cm$^{-1}$ result in negligible shock contribution and log($f_{\rm shocks}/f_{\rm star}$)>$-$2 for V$_{\rm exp}$ up to 100$-$120 km~s$^{-1}$. \\

Another commonly used way of investigating the photo- versus shock-excitation of astrophysical objects is through diagnostic diagrams such as \citet*{Sabbadin1977} or \citet{2006RMxAA..42...47R}, which are based on line ratios sensitive to these excitation processes. In Paper~I, we discussed these diagnostic tools, and given their relevance, our forthcoming work will analyse these and other diagnostic diagrams for a statistically significant sample of PNe that host low-ionization structures combined with predictions from photoionization and shock models.

\section{Conclusions}

The spectroscopic analysis of six PNe with embedded LISs was carried out. Physico-chemical properties of the PNe with different structural components were computed and compared with the results in the literature.

After analyzing the properties of the LISs and their host PNe, we reinforce the conclusions of previous works that the electron density of LISs are lower (or at most equal) than those of the PN main structures (rims and shells). 
Regarding electron temperatures, we point out a \textit{possible} different trends between estimations based on the nitrogen or the oxygen diagnostic lines. T$_e$[N~{\sc ii}] does not exhibit significant variations throughout the components, whereas T$_e$[O~{\sc iii}] appears to take slightly higher values for LISs relative to rims and shells, but also with much higher uncertainties, therefore not allowing robust conclusions.

In terms of abundances, as on Paper~I (and the previous ones in the series), we do not find 
any trend between components. We have to mention the lower O/H for a couple of LISs, which would thus imply a non-negligible variation. However, these LISs also have extremely uncertain (higher, sometimes exceeding 15000~K) T$_e$[O~{\sc iii}], which result in suspicious face values for their oxygen abundances.

In an attempt to better understand the excitation mechanisms present in LISs, the diagnostic diagram defined on Paper~I -- log($f_{\rm shocks}/f_{\rm star}$) -- was used, together with LISs kinematics. Considering that there are uncertainties on distances, pre-shock densities and radial velocities, we argue that the contribution of shocks to the excitation appears to be non-negligible
for the PNe in our sample. In particular, shocks seems to be unambiguously associated with Hen~2-186'~LISs, while for the remaining PNe the dependence on the inclination angle is much stronger, and since PNe do not have preferential orientations, shock excitation is not mandatory.

It seems evident the need to increase the sample of LISs to be able to perform a statistical analysis that takes into account all the properties found in recent years for these poorly understood structures. That is why these six PNe with LISs, jointly with a compilation of those already published in previous works, will be analysed in a forthcoming work. For this purpose, a statistical approach will be followed to derive the distribution of physical, chemical and main excitation mechanism of the low-ionization structures in planetary nebulae.

\section*{Acknowledgements}
We would like to thank Martín Guerrero, the referee, for his constructive comments and suggestions that helped to improve this work.
This paper contains data taken jointly with A. Mampaso and R. Corradi, back in 2001, during a wonderful week of observations at the Roque de los Muchachos. We are also as much as thankful to H. Schwarz (in memoriam), the observer of the Danish data. This research is support by a PhD grant from CAPES – the Brazilian Federal Agency for Support and Evaluation of Graduate Education within the Education Ministry. DGR acknowledges the CNPq grants 428330/2018-5 and 313016/2020-8. SA thanks the support under the grant 5077 financed by IAASARS/NOA.

\section*{Data Availability}

Some of the spectroscopic data underlying this article are available at the \hyperlink{http://casu.ast.cam.ac.uk/casuadc/ingarch/query}{Isaac Newton Group Archive}.


\typeout{}
\bibliographystyle{mnras}
\bibliography{mnras_P1} 







\onecolumn
\appendix
\begin{landscape}

\section{flux lines together with ionic and total abundances}
The flux emission lines measurements and obtained results are present in this section. The fluxes were corrected for interstellar reddening and then they were used to estimate T$_{e}$, N$_{e}$, ionic and total abundances for all nebulae present in this paper together with the smaller structures as rims, shells and LISs.

\vspace{0.5cm}

{\small

{\bf Table A1}: IC 4593

\begin{longtable}[h!]{cl|cccc|ccccc}

\hline\multicolumn{9}{r}{{Continued on next page}} \\ 
\endfoot

\endlastfoot

\hline
	  &		         &	\multicolumn{4}{c|}{PA=62}			    &  \multicolumn{5}{c}{PA=139}\\
\hline
  Line    & Ion                  & NE shell      & Neb         & SW Shell     & C	          & A	           & SE Shell    & Neb        & NW Shell       & B	     \\
	  & 	                 & 5px-3.5''     & 22px-15''    & 5px-3.5''    & 3px-2.1''        & 5px-3.5''      & 9px-6.3''   & 22px-15''    & 9px-6.3''      & 5px-3.5''  \\ 
  \hline
  3727.00 &   [O~{\sc ii}]b      & 55$\pm$1      & 45$\pm$1     & 54$\pm$2     & 126$\pm$4    & 151$\pm$17     & 42$\pm$2    & 44$\pm$4     & 38$\pm$6       & 74$\pm$8   \\
  3868.75 &[Ne~{\sc iii}]        & 28.3$\pm$0.4  & 33.3$\pm$0.8 & 23.2$\pm$0.9 & 22.7$\pm$1.7 & 31.5$\pm$10.3  &25.9$\pm$0.6 & 30.8$\pm$1.3 & 20.0$\pm$1.1   & 23.4$\pm$7.3   \\
  3889.05 & H~{\sc i}+He~{\sc i} & 20.4$\pm$0.3  & 14.3$\pm$0.8 & 19.2$\pm$0.8 & 21.2$\pm$1.3 & 24.5$\pm$12.3  &18.0$\pm$0.5 & 17.7$\pm$1.1 & 16.1$\pm$1.7   & 18.8$\pm$6.6   \\
  3967.46 &[Ne~{\sc iii}]        & 9.4$\pm$0.4   & 5.1$\pm$0.9  & 5.9$\pm$0.8  & 6.4$\pm$1.4  & 8.8$\pm$7.6    & 5.9$\pm$0.3 & 7.2$\pm$0.9  & 2.2$\pm$1.4    & 3.8$\pm$3.4   \\
  3970.07 & H~{\sc i}            & 16.8$\pm$0.4  & 16.2$\pm$0.9 & 16.2$\pm$0.8 & 15.8$\pm$1.5 & 17.0$\pm$7.8   &17.6$\pm$0.3 &16.2$\pm$0.9  & 16.5$\pm$1.4   & 17.3$\pm$6.3  \\
\end{longtable}


\newpage
{\bf Table A2}: Ionic and total abundances of IC 4593.

%
}
\hfill
\parbox{.45\linewidth}{
{\bf Table A4}: Ionic and total abundances of Hen 2-186.
}
\hfill
\parbox{.45\linewidth}{
{\bf Table A6}: Ionic and total abundances of Hen 2-429.

\newpage
{\bf Table A8}: Ionic and total abundances of NGC 3918.


{\bf Table A10}: Ionic and total abundances of NGC 6543.

%
}
\hfill
\parbox{.45\linewidth}{
{\bf Table 12}: Ionic and total abundances of NGC 6905.
\label{tab:n6905_2}
\end{longtable}
}

\end{landscape}

}


\bsp	
\label{lastpage}
\end{document}